\documentclass[11pt]{amsart}

\usepackage[hmargin=0.85in,twosideshift=0in,height=8.6in]{geometry}
\usepackage{amssymb,amsthm}
\usepackage{delarray,verbatim}
\usepackage{natbib}
\renewcommand{\cite}{\citet}

\usepackage{ifpdf}
\ifpdf
\usepackage[pdftex]{graphicx}
\else
\usepackage[dvips]{graphicx}
\fi

\usepackage{ifpdf}
\usepackage{color}
\definecolor{webgreen}{rgb}{0,.5,0}
\definecolor{webbrown}{rgb}{.8,0,0}
\definecolor{emphcolor}{rgb}{0.95,0.95,0.95}

\usepackage{hyperref}
\hypersetup{%
          colorlinks=true,
          linkcolor=webbrown,
          filecolor=webbrown,
          citecolor=webgreen,
          breaklinks=true}
\ifpdf
\hypersetup{pdftex,
            pdfstartview=FitH, 
            bookmarksopen=true,
            bookmarksnumbered=true
}
\else
\hypersetup{dvips}
\fi

\numberwithin{equation}{section} 

\newtheorem {thm}{Theorem}[section]
\newtheorem {prop}{Proposition}[section]
\newtheorem {lemm}[thm]{Lemma}

\newtheorem {cor}[thm]{Corollary}

\theoremstyle{remark}
\newtheorem {rem}{Remark}[section]

\DeclareMathOperator{\argmin}{argmin}

\newcommand{\ts}{\tilde{\tau}}
\newcommand{\eqd}{\stackrel{d}{=}}
\newcommand{\tP}{\tilde{\mathbb{P}}}

\newcommand{\R}{\mathbb R}
\newcommand{\N}{\mathbb N}

\newcommand{\PP}{\mathbb P}
\renewcommand{\P}{\mathbb P}
\newcommand{\A}{\mathcal A}

\newcommand{\E}{\mathbb E}

\renewcommand{\bar}{\overline}

\newcommand{\Fc}{\mathcal F}
\newcommand{\vP}{\vec{\Pi} }
\newcommand{\vp}{\vec{\pi} }
\newcommand{\vx}{\vec{x} }

\newcommand{\la}{\lambda}
\newcommand{\e}{\mathrm{e}}

\title{Optimal Trade Execution in Illiquid Markets}
\author{Erhan Bayraktar }
\address[E. Bayraktar]{Department of
  Mathematics, University of Michigan, Ann Arbor, MI 48109}
\email{erhan@umich.edu}
\thanks{E. Bayraktar is supported in part by the National Science Foundation. }

\author{Michael Ludkovski}
\address[M.\ Ludkovski]{Department of Statistics and Applied Probability, University of California Santa Barbara, CA 93106-3110\\ tel: (805)893-5634}
\email{ludkovski@pstat.ucsb.edu}

\begin{document}
 \begin{abstract}
 We study optimal trade execution strategies in financial markets with discrete order flow. The agent has a finite liquidation horizon and must minimize price impact given a random number of incoming trade counterparties. Assuming that the order flow $N$ is given by a Poisson process, we give a full analysis of the properties and computation of the optimal dynamic execution strategy. Extensions, whereby (a) $N$ is a fully-observed regime-switching Poisson process; and (b) $N$ is a Markov-modulated compound Poisson process driven by a hidden Markov chain, are also considered.
  We derive and compare the properties of the three cases and illustrate our results with computational examples.
 \end{abstract}


 \maketitle
\tableofcontents
\clearpage

\section{Introduction}
One of the most important problems faced by a stock trader is how to unwind large block orders of
security shares. Liquidation of a large position in a security is a challenge due to
two factors: (a) possible lack of a counterparty; and (b) price impact that depresses prices by increasing supply. This occurs because the immediate \emph{market resiliency} is limited and a single large order may exhaust all current buyers, bringing about dramatic price declines. Price impact implies that it is generally beneficial to split the order into several smaller blocks and sell each sub-block separately.  Presence of counterparties is less of a concern in traditional limit order book markets where a market maker is always quoting a price. However, trading in such markets may be disadvantageous due to information leak/privacy concerns. Indeed, by examining the order book, other participants may recognize the large trader and move against her, even if she attempts to split her trades. Thus, a recent trend involves trading in \emph{dark pool} markets where there is no order book and buyers/sellers are matched up electronically without revealing any information. Such dark trades minimize information leakage and dramatically reduce risk of adverse price movement compared to conventional limit book trading. However, liquidity becomes a major concern as there is no market-maker and no counterparty may be forthcoming. We refer to trade publications such as \cite{DB-newsletter} for more information on the evolving marketplace of dark pools and their numerous specification variations.

In this paper, we propose a new framework that explicitly takes into account such liquidity features of large order trades. Thus, we replace the classical continuous trading environment with a \emph{discrete} order book. In our model, incoming buy orders are represented by a Poisson process which encodes the order arrival times. To capture the empirical feature of splitting large orders into smaller pieces, we will focus on price impact and eschew consideration of actual prices. Larger trades involve a volume discount and therefore tend to carry higher spread versus the current quoted limit order price. Also, smaller trades are desirable in order to maintain anonymity and mitigate information leaks. Subject to the constraint that trades are only possible at order times, the objective of the agent is to execute her large order trade within a specified time-window while minimizing this price impact.

Most of the existing analysis of optimal execution has focused on limit order book markets, see e.g.\ \cite{schied2007,Almgren03,AlmgrenLorenz06,obi,SchiedSchoneborn08,SchiedSchonebornCARA}. Since a market maker is always present, all cited models assume a continuous-time trading environment, with the asset price usually represented by a diffusion price process. The price impact is decomposed into a temporary and permanent effects and execution strategies are specified in terms of \emph{liquidation rates} per unit time. The overall problem is then translated into a continuous or singular stochastic control formulation. Our approach is quite different, as in our case all trades are discrete and therefore an execution strategy corresponds to an impulse control setting. Also, in the above literature the optimal liquidation strategies turn out to be deterministic and can be sometimes explicitly determined. In contrast, our optimal strategies are intrinsically path-dependent and will be affected by the stochastic order flow. Finally, while the above papers typically consider an infinite horizon, we assume that the agent has a hard deadline to liquidate her large trade. Thus, time-to-maturity is a crucial variable in our setup and can be also used to express time-dependencies of real markets, where e.g.\ the opening and closing hours are typically much more liquid than midday. To sum up, our contribution is a new approach to modeling order execution liquidity in terms of point processes. As we show below, our models are flexible, allow for a quick implementation and admit fruitful probabilistic analysis.

Let us now outline the basic ingredients of our model. We assume that the order book is a Poisson process $N$ with arrival times $\sigma_i$ which denote the timestamp of the $i$-th order. In our base model we postulate that $N$ is a simple Poisson process with constant intensity $\lambda$ on a stochastic basis $(\Omega, \mathcal{F}, \PP)$. Suppose the agent has $k$ shares (or units) to sell and an execution horizon of $T$ time epochs. We postulate that at terminal date $T$ all unsold units are immediately disposed off as one large trade, e.g.\ through the traditional limit order book.  Thus, effectively there is always one more matching order arriving at $T$.  The price impact is represented in terms of a strictly increasing and strictly convex \emph{market depth} function $F$, where $F(a)$ represents the cost of placing a trade of size $a$ ($F(a)$ could also represent the average cost of a random price impact, assuming this randomness is independent of everything else in the model).

Let $\mathbb{F}=(\mathcal{F}_t)$, $\mathcal{F}_t = \sigma( N_s : 0 \le s \le t)$ be the filtration generated by the observation process.
Then the optimization problem of the agent can be written as
\begin{align}\label{eq:v-optim}
{v}(k,T) = \inf_{\xi \in {\A}_{k}} \E \left[ \sum_{i : \sigma_i \le T} F(\xi_{\sigma_i-}-\xi_{\sigma_i}) +F(\xi_T)\right]=: \inf_{\xi \in {\A}_{k}}v_{\xi}(k,T), \quad k \in \N_+, \,T \in \R_+,
\end{align}
where ${\A}_{k}$ is the set of all $\mathbb{F}$ -adapted, integer-valued, positive and non-increasing processes whose values change only at the time of jumps of the Poisson process $N$ with $\xi_0 = k$. The convexity of $F$ is interpreted as the limited market resiliency and encourages the agent to split the large $k$-order into smaller pieces. However, placing a smaller trade now is risky as no more orders might come in and the trader will be left with a large leftover at $T$ (which will carry a large associated penalty). Thus, the convexity of $F$ also represents the impatience of the agent in terms of current versus future trading and is formally similar to the risk-aversion level in \cite{SchiedSchoneborn08,SchiedSchonebornCARA}.

In terms of the stochastic control formulation, \eqref{eq:v-optim} is related to best choice problems with Poisson processes, see e.g.\
\cite{CowanZabczyk78,Bruss87}. In particular, \cite{Stadje87,Stadje90} studied a similar problem for a Poisson process in the context of multi-item dynamic pricing.
%

The mathematical problem in \eqref{eq:v-optim} is a compromise between a tractable analytical model and real markets. In general, the execution problem with illiquid trading is not so well-studied and a big challenge is to develop parsimonious models that will prescribe reasonable optimal
liquidation policies.  The use of a Poisson process for $N$ allows for a comprehensive analysis of \eqref{eq:v-optim} in Section \ref{sec:toy-model}, however, it is clearly not very rich to capture all the intricacies of real order books. Accordingly, we consider in Section \ref{sec:extend} several extensions to address such issues. Our base model allowed arbitrary trade sizes; in practice the agent is only able to trade up to the \emph{order size} which is the second dimension of the order flow. To reproduce this feature, in Section \ref{sec:constraints} we take $N$ to be a compound Poisson process, consisting of pairs $(\sigma_i,Y_i)$ of (order times, order sizes). Correspondingly, the original problem \eqref{eq:v-optim} is modified to constrain $\xi_{\sigma_i-}-\xi_{\sigma_i} \le Y_i$. Because of this constraint, the agent is expected to preemptively place larger orders in case a large matching order is forthcoming.

Second, the base model assumed that the intensity of $N$ was constant throughout the problem horizon.  Given widespread evidence that real markets experience different \emph{liquidity regimes}, in Section \ref{sec:regime-switching} we extend our model to the case where $N$ is a Markov-modulated Poisson process. Thus, we will assume $\lambda = \lambda(M_t)$ where $M$ is an (observed) independent Markov chain that describes the liquidity state of the market. Similarly, the distribution of sizes $Y_i$ will also be modulated by $M$.
In Section \ref{sec:partial-info} we then consider the even more realistic (and more complex) situation where traders do not observe $M$. Indeed, market participants do not know the current market liquidity and dynamically infer it given matched dark pool order flow. To capture this phenomenon, in Section \ref{sec:partial-info} we will assume that
the liquidity regime is modeled by a \emph{hidden} Markov chain $M$ that modulates the intensity and the jump distribution of crossing orders. To illustrate the different models mentioned above,
Section \ref{sec:examples} presents several computational examples; finally Section \ref{sec:conclusion} concludes and points possible future extensions.

\section{Analysis of the Optimal Liquidation Problem}\label{sec:toy-model}
In this section we analyze the properties of $v(k,T)$ as defined in \eqref{eq:v-optim}. The treatment below allows us to give a clear insight of the structure of $v(k,T)$ and leads to a particularly simple algorithm to compute $v$ and the associated optimal strategy, see Remark \ref{rem:computing-v}. Our first observation is that $v(k,T)$ satisfies the following dynamic programming equation:
\begin{equation}\label{eq:dpp}
v(k,T)=\E\left[ \min_{a \in \{1,\ldots,k\}} \left\{v(k-a,T-\sigma_1)+F(a)\right)\} \cdot 1_{\{\sigma_1 <
T\}}+F(k) \cdot 1_{\{\sigma_1 \geq T\}}\right].
\end{equation}
A more general version of this dynamic programming principle is proved in Proposition~\ref{prop:dpp}.

\subsection{Computing $v(k,T)$}
To illustrate the problem, let us explicitly compute $v(k,T)$ for a few values of $k$. First, for $k=1$ we trivially have  $v(1,T)=F(1)$, as one can simply wait till $T$ to make the single unit trade. Since $F$ is strictly convex, when there are two units to sell, it is clearly optimal to try to place two orders of size one. This will be possible as long as there is at least one arrival before date $T$, i.e.\ $N(T) \ge 1$ (recall that the remainder can always be disposed of at $T$). Applying \eqref{eq:dpp} and recalling the properties of the Poisson process yields $$v(2,T)=2F(1)\cdot(1-e^{-\lambda T})+F(2)\cdot e^{-\lambda T}.$$
When $k=3$, the agent needs to sell three units. Once an incoming order arrives, the agent should trade one unit, as getting rid of two or three units is not optimal (in the worse case, she will sell one unit now and the remaining two at $T$).  Conditioning on the time $\sigma_1$ of the first order, and using \eqref{eq:dpp} her expected minimal cost is then
\begin{align*}
v(3,T) & = F(3)\P(\sigma_1 > T) + \E[ (F(1) + v(2,T-\sigma_1)) 1_{\{\sigma_1 < T\}}] \\
& = F(3) e^{-\lambda T} + \int_0^T (F(1) + v(2, T-s))\lambda \e^{-\lambda s} \,ds \\
& = e^{-\lambda T} F(3)  + \lambda T e^{-\lambda T} F(2) + (3- 3e^{-\lambda T} - 2 \lambda T e^{-\lambda T}) F(1).
\end{align*}
The case $k=4$ is the first non-trivial case. Indeed, the agent can sell either one or two units when there is an incoming order (other strategies are clearly not optimal). This decision will be based on whether $ v(3,T-\sigma_1)+F(1) \geq v(2,T-\sigma_1)+F(2)$ at the first arrival time $\sigma_1$. If the latter inequality is true, then one is better off selling two units, otherwise a single unit is optimal to trade.  Observe that both sides of the last
inequality can be explicitly computed using previous formulas. From this computation it can be observed that as time to
maturity, $T$, becomes smaller the agent gets more impatient and trades two units instead of
one as soon as there is an arrival. Thus, there exists a critical threshold $t^{(4,2)}$ such
that if $T-\sigma_1 \ge t^{(4,2)}$ then it is optimal to trade just one unit, and if $T-\sigma_1
< t^{(4,2)}$ then it is optimal to trade two units.

Let $a(k,T)$ be the optimal order size to place when an order arrives given that one has $k$ units remaining and $T$ epochs until the terminal date. Then the above analysis shows that $a(1,T) = a(2,T) = a(3,T) = 1$ for all $T\ge 0$, while $a(4,T) = 1 + 1_{\{T< t^{(4,2)}\}}$. In general, it follows from \eqref{eq:dpp}  that
\begin{align}\label{eq:a-defn}
a(k,T) = \argmin_{a \in \{1,\ldots,k\}} \{ v(k-a,T)+F(a) \}.
\end{align}
The above equation is simply the dynamic programming principle that says that the best \emph{immediate} action is to sell $a$ units, such that the sum of the current cost $F(a)$ and expected future costs as represented by the value function $v(k-a,T)$ is minimized. To avoid ambiguity, we will assume that if the minimizer in \eqref{eq:a-defn} is not unique, then $a(k,T)$ is the \emph{smallest} minimizer.

We conclude this section with upper and lower bounds for $v(k,T)$. The next lemma gives an easy to compute lower bound for the value function. Below, we extend the domain of $F$ to the whole positive real line such that $F :\R_+ \to \R_+$ is still strictly convex and increasing.
\begin{lemm}\label{lem:lb} We have
\begin{equation}\label{eq:genie}
v(k,T) \geq \sum_{n<k-1} F\left(k \over n+1\right)\P(N(T)=n)+k F(1)\P(N(T) \geq k-1) := \underline{v}(k,T).
\end{equation}
\end{lemm}

\begin{proof}
Consider a genie who is affected by the randomness but for each state of the world can tell how many arrivals there will be. Let us assume also that the genie is allowed to divide up her orders into non-integral bits of size $\ge 1$. Then, conditional on knowing $N(T)=n < k-1$, the genie should execute $n+1$ trades of size $k/(n+1)$ (the last trade comes at the period close). Consequently, the right hand side of \eqref{eq:genie} is the genie's solution to \eqref{eq:v-optim} which is clearly better than the optimal solution of the mortal, who does not possess any clairvoyance about $N$ and can only divide up her blocks into integral units.
\end{proof}

As counterpart to Lemma \ref{lem:lb}, we have the following tight upper bound to the value function.

\begin{lemm}\label{lem:ub}
\begin{multline}\label{eq:dumm}
v(k,T) \leq \min_{c \in \N_+}  \bar{v}^c(k,T) \triangleq \min_{c \in \N_+}  \Bigl\{\left(\Big\lfloor \frac{k}{c}\Big\rfloor \cdot F(c)+F\left(k-c \cdot \Big\lfloor \frac{k}{c}\Big\rfloor \right)\right)\cdot\P\left(N(T)\geq \Big\lfloor \frac{k}{c}\Big\rfloor \right)\\ +\sum_{ n = 0}^{\big\lfloor \frac{k}{c} \big\rfloor - 1}\left(n \cdot F(c) +F\left(k-n \Big\lfloor \frac{k}{ c}\Big\rfloor \right)\right)\cdot \P(N(T)=n)\Bigr\},
\end{multline}
in which $\lfloor x/ c\rfloor $ is the largest integer smaller than $x/c$.
\end{lemm}

\begin{proof}
The right hand side of \eqref{eq:dumm} is the cost of a constant $c$-strategy. This is the strategy where the agent insists on trading $c$ units at each arrival time until terminal date $T$, whence the remainder is liquidated. Although she originally optimizes over $c$, clearly this is a sub-optimal strategy.
The bound in  \eqref{eq:dumm} becomes tight as $T\to \infty$, the liquidity risk vanishes and the optimal strategy is to always trade a single unit $c^* =1$.
\end{proof}

\subsection{Properties of the value function}\label{eq:prop-v}
We now present a series of Lemmas that describe the properties of $v$ and $a$. This section then culminates with Proposition \ref{prop:1} which summarizes our analysis below.

In parallel with the original formulation in \eqref{eq:v-optim} in terms of dynamic controls $\xi$, one may also describe Markov control strategies as  $\{b(k,T) : k \in \N, T \in \R_+ \}$, specifying the trading amount conditional on still having $k$ units left with time horizon of $T$ periods. Given such $\{b(k,T)\}$, the corresponding dynamic unit inventory process is denoted by $\xi^{(b,k,T)}_t$ and satisfies
\begin{equation}\label{eq:cont-dyn}
d \xi^{(b,k,T)}_t=-b\left(\xi^{(b,k,T)}_t,T-t\right)dN_t, \quad \xi^{(b,k,T)}_0=k.
\end{equation}
Economically, $\xi^{(b,k,T)}_t$ represents the remaining number of units at date $t$ when employing the execution strategy $b$. Using $a$ to denote the strategy characterized by \eqref{eq:a-defn}, it follows that an optimal inventory process for \eqref{eq:v-optim} is given by $\xi^* \equiv \xi^{(a,k,T)}$. In particular, an optimal control is of the Markovian feedback type.

The following lemma immediately follows from the definition of the value function in \eqref{eq:v-optim}.

\begin{lemm}\label{lem:vwrt-T}
The function $k \to v(k,T)$ is increasing and the function $T \to v(k,T)$ is decreasing.
\end{lemm}

\begin{proof}
The above results are model-free in the sense that they depend solely on the convexity of $F$ and not on any properties of the arrival process $N$. Thus, it is instructive to give a short proof.
Let $\xi$ be any admissible control for $v(k,T)$. Then $\xi$ is also admissible for $v(\ell,T)$ for any $\ell \ge k$, which immediately establishes the first part of the lemma. Moreover, for any $T' > T$, define a control $\xi'$ via $\xi_t' = \xi_t$ for $t \le T$ and $\xi_{\sigma_i-}'-\xi_{\sigma_i}' = 1_{\xi_{\sigma_i}'>0}$ for $T < \sigma_i \le T'$. Then $\xi'$ is an admissible control for $v(k,T')$. Moreover, due to the convexity of $F$, the pathwise cost of $\xi'$ is less than or equal to the pathwise cost of $\xi$,
$$
\sum_{i : \sigma_i \le T} F(\xi_{\sigma_i-}'-\xi_{\sigma_i}') + \sum_{j: T < \sigma_{j} \le T} F(1) + F(\xi_{T'}') \le \sum_{i : \sigma_i \le T} F(\xi_{\sigma_i-}-\xi_{\sigma_i}) + F(\xi_T) \quad \PP-\text{a.s.},
$$
with strict inequality if $\xi_{T} > 1$ and $N(T')-N(T) > 0$. It follows that $v(k,T'; \xi') \le v(k,T,\xi)$ with strict inequality as long as $\PP( N(T) = 0, N(T')-N(T) > 0)>0$. Note that the last statement is satisfied for $N$ a Poisson process and any $T<T'$.
\end{proof}

The following basic lemma shows that the slope of $v$ is smaller than that of $F$. 

\begin{lemm}\label{eq:F-v}
For any $k_1 > k_2$ and $t$ we have $v(k_1,T) - v(k_2,T) < F(k_1) - F(k_2)$. Alternatively, $F(k)-v(k,T)$ is increasing in $k$.
\end{lemm}
\begin{proof}
Let $\xi^{k_2}$ denote $\xi^{(a,k_2,T)}$. Recall that $v_\xi(k,T)$ denotes the expected performance of any control $\xi$. Interpreting $\xi^{k_2}$ as a sub-optimal control for $v(k_1,T)$ (which disposes of the extra $k_1-k_2$ units at maturity), we have
\begin{align*}
v(k_1,T) - v(k_2,T) & \le v_{\xi^{k_2}}(k_1,T) - v_{\xi^{k_2}}(k_2,T) \\
& = \E\left[ \sum_{i=0}^{k_2} (F(i+k_1-k_2) - F(i))1_{\{\xi^{k_2}_T= i\}} \right]\\
& < \sum_{i=0}^{k_2} \E \Bigl[ (F(k_1) - F(k_2))1_{\{\xi^{k_2}_T = i\}} \Bigr] = F(k_1) - F(k_2),
\end{align*}
where the second inequality follows from the convexity of $F$, whereby $F(a+y)-F(y)$ is
increasing in $y$.
\end{proof}



The following lemma shows that if one starts with more units initially and sells them in an
optimal way, then one will always have more units at any later point in time (an intuitive
observation).
\begin{lemm}\label{lem:inv-proc}
Let $\xi^{k}_t$ denote $\xi^{(a,k,T)}_t$, $k \in \N_+$. Then for $\ell \geq k$ we have that  $\xi^{\ell}_t \geq \xi^k_t$ for all $ t \in [0,T]$.
\end{lemm}

\begin{proof}
First note that if at any date $s \le t$ we would have $\xi^{\ell}_s=\xi_s^k$, then it follows from \eqref{eq:cont-dyn} and the Markov nature of $a(k,T)$ that
 for all $s' \ge s$ we will have $\xi^{\ell}_{s'}=\xi_{s'}^k$ as well. Thus, to have $\xi_s^{\ell}<
\xi_s^k$ on a set $A$ of strictly positive probability there necessarily must be an arrival $\sigma_j$ such that $d_l:=\xi^{\ell}_{\sigma_j-}>
\xi^k_{\sigma_j-} := d_k$ and $b_\ell := \xi^{ \ell}_{\sigma_j}<
\xi^k_{\sigma_j} := b_k$ on  $A$.
By construction, $b_\ell = d_\ell - a(d_\ell, T-\sigma_j)$ and $b_k = d_k - a(d_k, T-\sigma_j)$. Moreover,
\begin{align}\label{eq:opt-a}
\left\{\begin{aligned} a(d_\ell,T-\sigma_j) & =: a_\ell = \argmin_a \{ v(d_\ell-a,T-\sigma_j)+F(a) \}; \\
a(d_k,T-\sigma_j) &=: a_k = \argmin_a \{ v(d_k-a,T-\sigma_j)+F(a) \}. \end{aligned}\right.
\end{align}
 Define $c_\ell = d_\ell-d_k+a_k > a_k$, and $c_k=d_k-d_\ell+a_\ell<a_\ell$.
Therefore from \eqref{eq:opt-a} (and recalling that $a_\ell$ is the smallest minimizer, while $a_\ell > c_\ell$)
$$ \left\{\begin{aligned}  v(d_\ell -a_\ell, T-\sigma_j) + F(a_\ell) & < v(d_\ell-c_\ell,T-\sigma_j)+F(c_\ell),
\\
v(d_k-a_k,T-\sigma_j)+F(a_k) & \le v(d_k-c_k,T-\sigma_j)+F(c_k). \end{aligned}\right.$$ Re-arranging, we obtain
\begin{align}\label{eq:contradiction}
\left\{\begin{aligned}  v(d_\ell-c_\ell,T-\sigma_j) -v(d_\ell-a_\ell,T-\sigma_j) & > F(a_\ell)-F(c_\ell), \\
v(d_k-a_k,T-\sigma_j) -v(d_k-c_k,T-\sigma_j) & \le F(c_k)-F(a_k). \end{aligned}\right.
\end{align}
However, the left-hand-sides of both equations in \eqref{eq:contradiction} are the same by construction and are in fact
equal to $v(b_k,T-\sigma_j)-v(b_\ell,T-\sigma_j)>0$. On the other hand, since $a_\ell > c_k$ and
$c_\ell>a_k$, while $$a_\ell-c_\ell = c_k-a_k = a_\ell-a_k+d_k-d_\ell = b_k - b_\ell > 0,$$
by the convexity of $F$ we must have $F(a_\ell)-F(c_\ell) \ge F(c_k)-F(a_k),$ contradicting
\eqref{eq:contradiction}.
\end{proof}

The above lemma implies the following useful corollary regarding optimal actions for different inventory levels.
\begin{cor}\label{cor:a-diff}
For any $T \in \R_+$ and $\ell \leq  k$, we have $a(k,T)-a(\ell,T) \le k-\ell $ for all $t\ge 0$. In particular, $a(k+1,T) \le a(k,T)+1$.
\end{cor}
Corollary \ref{cor:a-diff} follows from Lemma \ref{lem:inv-proc} since the given relation between optimal actions is necessary to keep the corresponding inventory processes ordered correctly.

\begin{lemm}\label{lem:main-v-conv}
We have $v(k,T)$ is ``convex'' in $k$, that is for any $k \in \N_+$
\begin{equation} \label{eq:convex}
v(k,T)-v(\ell,T) \geq v(k-n,T)-v(\ell-n,T), \quad \forall\ell \in \{1,\cdots,k\}, \forall n \in \{1,\cdots,l\}.
\end{equation}
Also,
for any $T \in \R_+$ and $\ell  \leq k$,
\begin{align}\label{eq:a-incr}
a(\ell, T) \leq a(k,T).
\end{align}

\end{lemm}

\begin{proof}
We will prove both of the above statements together by induction. Note that \eqref{eq:convex} holds when $k=1$ since $v(0,T)=0$. Also $a(1,T) \geq a(0,T)=0$.
Suppose that
\eqref{eq:convex} and \eqref{eq:a-incr} hold for some $k \geq 1$. We will show that they are
also true when $k$ is replaced by $k+1$.
It is enough to prove that
\begin{equation}\label{eq:v-induc}
v(k+1,T)-v(k,T) \ge v(k,T)-v(k-1,T),
\end{equation}
and that $a(k+1,T) \geq a(k,T)$.

First, by definition $a(k+1,T) = \argmin_a \{ v(k+1-a,T) + F(a) \}$. Now suppose that $a(k,T) >
b \ge 1$. This implies that
\begin{align*}
v(k-b,T)+F(b) &> v(k-a(k,T),T) + F(a(k,T)) \\
\Longleftrightarrow \quad v(k-b,T)-v(k-a(k,T),T) &> F(a(k,T))-F(b) \\
\Longrightarrow \quad v(k+1-b,T)-v(k+1-a(k,T),T) &> F(a(k,T))-F(b)>0\\
\Longrightarrow \quad a(k+1,T) & \neq b,
\end{align*}
since the sale of $b$ shares is less preferable than selling $a(k,T)$ shares.
The third line follows from the induction hypothesis since $k+1-b \leq k$. Since $a(k+1,T) \neq b$ for any $b<a(k,T)$ we necessarily have that $a(k+1,T) \geq a(k,T)$.

Thanks to the fact that  $a(k+1,T) \geq a(k,T)$ for all $T \in \R_+$, the induction hypothesis on $a$, and the dynamics of $\xi^i \equiv \xi^{(a,i,T)}$ given in \eqref{eq:cont-dyn}
  we have that $\xi^{k+1}_{\sigma_{n}-}-\xi^{k+1}_{\sigma_{n}} = \xi^k_{\sigma_{n}-}-\xi^k_{\sigma_{n}}+\Delta_{\sigma_{n}}$, where $\Delta_{\sigma_{n}} \in \{0,1\}$ ($\Delta_{\sigma_{n}} \leq 1$ due to Corollary \ref{cor:a-diff}).
 The process $\Delta$ should be thought of as the ``additional'' action needed to sell one \emph{more} unit starting with $k$ units. Now, the left-hand-side of \eqref{eq:v-induc} becomes
$v(k+1,T) -v(k,T)  =$
\begin{align}\notag
 &= \E \left[ \sum_{n : \sigma_n \le T} 1_{\{\Delta_{\sigma_{n}} >
0\}}\{F(\xi^{k+1}_{\sigma_{n}-}-\xi^{k+1}_{\sigma_{n}} )-F(\xi^k_{\sigma_{n}-}-\xi^k_{\sigma_{n}})\} +1_{\{\sum_n \Delta_{\sigma_{n}} =
0\}}\{F(\xi_T^{k+1})-F(\xi^k_T)\}\right] \\ \label{eq:v-diff}
& = \E \left[ \sum_{n : \sigma_n \le T} 1_{\{\Delta_{\sigma_{n}} >
0\}}\{F(\xi^k_{\sigma_{n}-}-\xi^k_{\sigma_{n}})+1)-F(\xi^k_{\sigma_{n}-}-\xi^k_{\sigma_{n}}))\}
+1_{\{\sum_n \Delta_{\sigma_{n}} =
0\}}\{F(\xi_T^{k}+1)-F(\xi^k_T)\}
\right].
\end{align}
Let us analyze the right-hand-side of \eqref{eq:v-induc} .
Define the control $\xi'$ by $\xi'_0=k$ and
$\xi'_{\sigma_n-}-\xi'_{\sigma_n}=\xi^{k-1}_{\sigma_n-}-\xi^{k-1}_{\sigma_n}+\Delta_{\sigma_{n}}$.
This is an admissible control for
selling $k$ units. Then,
\begin{align*}
v&(k,T)-v(k-1,T)  \le \E \left[ \sum_{n : \sigma_n \le T} F(\xi'_{\sigma_n-}-\xi'_{\sigma_n})+F(\xi'_T) \right] - \E\left[ \sum_{n : \sigma_n \le T} F(\xi^{k-1}_{\sigma_n-}-\xi^{k-1}_{\sigma_n})+F(\xi^k_T)\right] \\
&= \E \left[ \sum_{n: \sigma_n \le T} 1_{\{\Delta_{\sigma_{n}} >
0\}}\{F(\xi^{k-1}_{\sigma_n-}-\xi^{k-1}_{\sigma_n}+1)-F(\xi^{k-1}_{\sigma_n-}-\xi^{k-1}_{\sigma_n})\} +1_{\{\sum_n \Delta_{\sigma_{n}} =
0\}}\{F(\xi_T^{k-1}+1)-F(\xi^{k-1}_T)\}\right] \\
& \le \E \left[ \sum_{n : \sigma_n \le T} 1_{\{\Delta_{\sigma_{n}}>
0\}}\{F(\xi^{k}_{\sigma_n-}-\xi^{k}_{\sigma_n}+1)-F(\xi^{k}_{\sigma_n-}-\xi^{k}_{\sigma_n})\} +1_{\{\sum_n \Delta_{\sigma_{n}} =
0\}}\{F(\xi_T^{k}+1)-F(\xi^{k}_T)\}\right] \\
& = v(k+1,T)-v(k,T).
\end{align*}
The last inequality is by the convexity of $F$ and the induction hypothesis on $a$ from which it follows that $\xi^{k-1}_{\sigma_n-}-\xi^{k-1}_{\sigma_n} \leq \xi^{k}_{\sigma_n-}-\xi^{k}_{\sigma_n}$. The last equality is from \eqref{eq:v-diff}. This completes the proof.
\end{proof}

To better connect Lemma \ref{lem:main-v-conv} with the notion of convexity, we state the following corollary:
\begin{cor}\label{cor:convex}
Fix $a,k \in \mathbb{N}$ with $a< k$. Then for any $b \in \N_+$ with $a<b \leq k$ we have that
\begin{equation}\label{eq:conve}
v(k-a-1,T) \leq \alpha v(k-b,T)+(1-\alpha)v(k-a,T),
\end{equation}
in which $\alpha=1/(b-a)$.
\end{cor}

\begin{proof}
We will prove this statement by induction.
Note that \eqref{eq:conve} or equivalently,
\begin{equation}\label{eq:conintepf}
v(k-a,T)-v(k-b,T) \leq (b-a) [v(k-a,T)-v(k-a-1,T)]
\end{equation}
holds for $b=a+1$.
Let us assume that \eqref{eq:conintepf} holds for $b=a+n$ (in which $n$ is such that $a+n+1 \leq k$), i.e.,
\[
v(k-a,T)-v(k-a-n,T)\leq n [v(k-a,T)-v(k-a-1,T)].
\]
On the other hand,
\[
v(k-a-n,T)-v(k-a-n-1,T) \leq v(k-a,T)- v(k-a-1,T),
\]
thanks to Lemma~\ref{lem:main-v-conv}. Adding the last two inequalities, we obtain \eqref{eq:conintepf}
for $b=a+n+1$.
\end{proof}
The above corollary in particular implies that there are at most two minimizers in \eqref{eq:a-defn}. Indeed,
%
%
if $a_1 = a(k,T) < a_2 = a_2(k,T)$ are both minimizers in \eqref{eq:a-defn}, i.e.
\begin{equation}\label{eq:twmin}
v(k-a_1,T)+F(a_1)=v(k-a_2,T)+F(a_2),
\end{equation}
then with $\alpha={1 \over a_2-a_1}<1$, we obtain
\begin{align*}
v(k-a_1-1)+F(a_1+1) & \le \alpha v(k-a_2,T)+(1-\alpha)v(k-a_1,T) + \alpha F(a_2)+(1-\alpha)F(a_1)\\
& \le v(k-a_1,T)+F(a_1),
\end{align*}
where the last line used \eqref{eq:twmin}. This is a contradiction as $a(k,T)=a_1$ is the smallest minimizer of \eqref{eq:a-defn}.

%
%
\begin{lemm}\label{lemm:G}
Define \begin{align}\label{defn:G}
 G(k,T) \triangleq v(k,T) -\min_{a\in \{0,1,\ldots,k\}} [v(k-a,T) + F(a)].
\end{align}
The map $k \to G(k,T)$ is non-decreasing for all $T \in \R_+$.
\end{lemm}

\begin{proof}
We will show that $G(k,T) \geq G(\ell,T)$ for $k \geq \ell.$
Since $a(\ell,T) \leq a(k,T),$
\begin{align*}
G(k,T) & \geq v(k,T)-(v(k-a(\ell,T),T)+F(a(\ell,T))) \\ & \geq v(\ell,T)-(v(\ell-a(\ell,T),T)+F(a(\ell,T))) =G(\ell,T),
\end{align*}
in which the second inequality follows from Lemma~\ref{lem:main-v-conv}.
\end{proof}

The quantity
$G(k,T)$ in Lemma~\ref{lemm:G} represents the maximal gain from an immediate impending trade. Lemma~\ref{lemm:G} has the interpretation that the more units the agent still has, the more eager she is to sell them and so the benefit of a matching order is larger. The next lemma shows that $G$ is also related to the time-derivative of $v$.

\begin{lemm}\label{lem:v-deriv}
The derivative of $v$ with respect to time-to-maturity is
\begin{equation}\label{eq:v-deriv}
\partial_T v(k,T) = -\lambda G(k,T).
\end{equation}
\end{lemm}
\begin{proof}

For $h>0$, let $A=\{\sigma_1>h\}$, $B = \{ \sigma_1 < h, \sigma_2>h \}$ and $C=(A\cup B)^c$.
We have that $\P(A)=\e^{-\lambda h}$, $\P(B)=\lambda h \e^{-\lambda h}$ and $\P(C)=o(h)$.
Using the dynamic programming principle, we can write
\[
 v(k,T+h) = \E[ v(k,T) 1_{A} + (v(k,T)-G(k,T)) 1_{B} + X 1_{C}]
\]
in which $X$ is a bounded random variable. Then sending $h \to 0$ we obtain
\begin{align*}
\lim_{h\to 0} \frac{ v(k,T+h)-v(k,T) }{h} & = \lim_{h\to 0}  \frac{ \E[ v(k,T)(1_{A \cup B}) -
G(k,T) 1_{B}]-v(k,T) + o(h)}{h} \\
& = \lim_{h\to 0}  \frac{-\lambda h G(k,T) + o(h)}{h} = -\lambda G(k,T).
\end{align*}
\end{proof}
Using Lemma \ref{lem:v-deriv} we may complete our description of the properties of  $a(k,T)$.  First, the next lemma shows that optimal trading amount decreases as the horizon becomes longer.
\begin{lemm}\label{lem:awrtT}
For any $S>T$, $a(k,S) \leq a(k,T)$, $ \forall k \in \N_+$.
\end{lemm}

\begin{proof}
 For any $b > a(k,T)$
\begin{align*}
 v(k - b, T) + F(b) & > v(k-a(k,T),T)+F(a(k,T)) \\ \Longleftrightarrow \qquad v(k-a(k,T),T)-v(k-b,T) & < F(b)-F(a(k,T)).
 \end{align*}
 We have that $\partial_T v(k,T) \le \partial_T v(\ell,T)$ for $\ell \leq  k$, due to Lemmas~\ref{lemm:G} and \ref{lem:v-deriv}. Therefore,
 \[
v(k-a(k,T),S)-v(k-b,S) \le v(k-a(k,T),T)-v(k-b,T) < F(b)-F(a(k,T))
 \]
which implies that $a(k,T)$ performs strictly better than action $b$ for the minimization problem $\min_{a \in \{0,\cdots l\}} \{v(k-a,S)+F(a)\}$ which implies that $b \neq a(k,S)$, which is the smallest minimizer for this problem. Since this holds for any  $b > a(k,T)$ we necessarily have that $a(k,T) \geq a(k,S)$.
\end{proof}
In the next lemma we shall see that $T \to a(k,T)$ decreases to 1.
\begin{lemm}
$\lim_{T \to \infty} v(k,T)=k F(1)$ and $\lim_{T \to \infty} a(k,T)=1$. We also have that $a(k,0)=\lfloor k/2\rfloor$.
\end{lemm}

\begin{proof}
Recall from Lemma \ref{lem:ub} that $v(k,T) \le \bar{v}^1(k,T)$ where $\bar{v}^1$ denotes the performance of a constant 1-strategy that always sells a single unit. Since
$$
\bar{v}^1(k,T) = k F(1) \PP(N(T) \ge k)+\sum_{n=0}^{k-1} (n F(k)+F(k-n)) \PP(N(T)=n) \to k F(1) \quad \text{as  }T\to\infty,
$$
while $v(k,T) \ge k F(1)$ $\forall T$, the first statement of the lemma follows.

Let us choose a positive $0<\delta<F(2)-2F(1)$.  Fix $k>0$; by above, for large enough $\mathcal{T}$, we have that $v(a,T) \leq a F(1)+\delta$ for all $a \in \{1,\cdots, k\}$. Then by convexity of $F$
$$
v(k-1,T)+F(1) \le (k-1)F(1) +\delta + F(1) < (k-c)F(1)+F(c) \le v(k-c,T)+F(c),
$$
for any $T \geq \mathcal{T}$ and any $1 < c < k$. 
Comparing with the definition of $a(k,T)$ in \eqref{eq:a-defn}, we conclude that $a(k,T)=1$ for $T \geq \mathcal{T}$.
\end{proof}

\begin{cor}\label{cor:ajsb1}
There exist distinct thresholds $t^{(k,i)}$ such that $a(k,T)=i$ when
\begin{equation}\label{eq:jumpth}
t^{(k, i+1)} <  T\leq t^{(k, i)}.
\end{equation}
\end{cor}
\begin{proof}
The basic idea of the corollary follows from Lemma~\ref{lem:awrtT}. It remains to show that the thresholds are distinct, i.e.\ $t^{(k,i)} < t^{(k,i-1)}$, so that as a function of $T$, $a(k,T)$ experiences jumps of size 1 only.

Toward a contradiction, suppose that there exists $T$ and level $k$ such that $a(k,T-)-a(k,T) > 1$. Let $a=a(k,T)$ and $b=a(k,T-)> a+1$.  Since $1 \le a(k,\cdot) \le \lfloor k/2\rfloor$ is non-increasing and has at most $\lfloor k/2\rfloor-1$ jumps,  there exists $\delta > 0$ such that $b=a(k,T-s)$ for all $s< \delta$. By optimality of $b$ we have that
\begin{align*}
  v(k-b,T-s) +F(b) = v(k-a(k,T-s),T-s)+F(a(k,T-s)) & \le v(k-a,T-s)+F(a) \quad\forall s < \delta 
\end{align*}
Therefore, by continuity of the value function in $T$, and optimality of $a$ at $T$ we must have
\begin{align}\label{eq:a-jumps-by-one}
v(k-a,T)+F(a) = v(k-b,T)+F(b).
\end{align}
Let $\alpha = 1/(b-a) \in (0,1)$.
By the strict convexity of $F$ we have that $F(a+1) < \alpha F(b) + (1-\alpha)F(a)$. Similarly, by Corollary~\ref{cor:convex}, we have that $v(k-a-1,T) \le \alpha v(k-b,T) + (1-\alpha)v(k-a,T)$. Adding the two latter equations together we obtain
\begin{align*}
v(k-a-1,T)+F(a+1) & < \alpha( v(k-b,T)+F(b) ) + (1-\alpha)(v(k-a,T)+F(a)) \\
& = v(k-a,T)+F(a), \qquad\qquad \text{by \eqref{eq:a-jumps-by-one}},
\end{align*}
which contradicts the optimality of $a$.
\end{proof}
As a corollary of Lemma~\ref{lem:v-deriv} and Corollary~\ref{cor:ajsb1} we have the following result.\begin{cor}
The function $T \to v(k,T)$ is decreasing and convex. The second derivative of $v$ with respect to $T$ is continuous except at $T \in \{ t^{(k,i)} : i=1,\ldots, \lfloor k/2 \rfloor \}$ (see \eqref{eq:jumpth}).
\end{cor}

\begin{proof}
We already know that $v(\cdot,T)$ is decreasing from Lemma \ref{lem:vwrt-T}.
For any $T \neq t^{(k,i)}$ we have from combining \eqref{defn:G} and \eqref{eq:v-deriv} that
 \begin{equation}\label{eq:derG}
 \partial_T G(k,T) = -\lambda (G(k,T) - G(k-a(k,T),T)),
 \end{equation}
since $a(k,T)$ is constant in a neighborhood of $T$ thanks to Corollary~\ref{cor:ajsb1}.
 When $T=t^{(k,i)}$, the right derivative of $G$ is still equal to \eqref{eq:derG} since $T \to a(k,T)$ is right continuous. But the left derivative is equal to  $\partial_T G(k,T-) = -\lambda (G(k,T) - G(k-a(k,T)-1,T))$. Recalling  Lemma~\ref{lem:v-deriv} we see that the second derivative of $v$ with respect to $T$ has a discontinuity at  $T=t^{(k,i)}$.

On the other hand, by Lemma \ref{lemm:G}, derivatives of $G$ with respect to $T$ are negative and thus the second derivative of $v$ is positive.
\end{proof}

Another corollary of Lemma \ref{lem:awrtT} is the effect of the arrival intensity $\lambda$ of $N$.
\begin{cor}\label{lem:a-lambda-dep}
The value function $v(k,T)$ and optimal action $a(k,T)$ are decreasing in $\lambda$.
\end{cor}

Note that we have the scaling property
$v(k,T; \lambda) = v(k, \alpha T; \lambda/\alpha)$ for any $\alpha > 0$ since the main parameter is intensity of arrivals per effective horizon. Thus, dependence of $v$ (and $a$) on $\lambda$ is equivalent to its inverse dependence on time horizon. Below we give a second proof using the concept of coupling. This approach will be re-used later in Section \ref{sec:partial-info}.

\begin{proof}
Consider two Poisson processes $N_1$, $N_2$ with intensities $\lambda_1 > \lambda_2$. Then one may
construct a probability space $(\Omega', \mathcal{F}', \PP')$ and random variables $\tau^{(i)}_k$, $i=1,2$, $k=1,2,\ldots$ such that $\tau^{(i)}_k \sim \mathcal{E}xp(\lambda_i)$ and $\tau^{(1)}_k \le \tau^{(2)}_k$ $\PP'$-almost surely. Letting $N_i'(t) = \max( k : \sum_{j=1}^k \tau^{(i)}_j \le t)$ we obtain two coupled copies $N_1', N_2'$ of $N_1$, $N_2$, such that $\PP'( N_1'(t) \ge N_2'(t) \;\forall t) = 1$. Now it is fairly obvious that $v^{\lambda_1}(k,T) \le v^{\lambda_2}(k,T)$ since working under $\PP'$, the first case has almost surely more arrivals than
the second case. Formally, let us define a deterministic time-change by $\tau(t) = \lambda_1/\lambda_2 t$. Then $\PP'(\tau^{(1)}_k \in dt) = \PP'(\tau^{(2)}_k \in \tau(dt))$, which implies $\PP'(N_1'(t) \le j) = \PP(N_2'(\tau(t)) \le j)$ for all $j$ and therefore
$v^{\lambda_1}(k,T) = v^{\lambda_2}(k,\tau(T))$ (map any control $\xi$ for $v^{\lambda_1}(k,T)$ into a control $\xi_{\tau(t)}$ for  $v^{\lambda_2}(k,\tau(T))$). Now, since $\tau(T) > T$ it follows that $v^{\lambda_2}(k,T) > v^{\lambda_2}(k,\tau(T)) = v^{\lambda_1}(k,T)$.
\end{proof}

The following Proposition is the main result of this section and summarizes all the above analysis.
\begin{prop}\label{prop:1}
Consider the problem
$$
v(k,T) = \inf_{\xi \in \A_{k}} \E \left[ \sum_{i : \sigma_i \le T} F(\xi_{\sigma_i-}-\xi_{\sigma_i}) +F(\xi_T)\right].
$$
Then the optimal strategy is given by \eqref{eq:a-defn} and:
\begin{enumerate}
\item $k \to v(k,T)$ is non-decreasing, ``convex", and $v(k+1,T)-v(k,T) < F(k+1) -F(k)$ for all $k,T$.

\item $T \to v(k,T)$ is decreasing and convex. Moreover, $T \to \partial_{T}^2 v(k,T)$ is discontinuous only at at $T=t^{(k,i)}$ (see \eqref{eq:jumpth}).

\item $\partial_T v(k,T) = -\lambda (v(k,T) -\min_{a\in \{0,1,\ldots,k\}} [v(k-a,T) + F(a)]) <0$. Moreover $\partial_T v(k,T)$ is increasing in $k$.

\item $k \to a(k, T)$ is non-decreasing and increases by jumps of size 1 only.

\item $T  \to a(k,T)$ is non-increasing and right continuous with $a(k,0)=\lfloor k/2\rfloor$ and $\lim_{T \to \infty}a(k,T)=1$. Moreover, its jumps are of size 1. The jumps occur at  $T=t^{(k,i)}$.
\end{enumerate}
\end{prop}

\begin{rem}\label{rem:computing-v} \textbf{A word on the computation of the value function and the optimal action.}
Using the above results, one may easily compute $v(k,T)$ for any depth function $F(\cdot)$ by using the coupled family of first-order ordinary differential equations \eqref{eq:v-deriv} over a time grid.
  Note that given $v(k,T), a(k,T)$, finding the minimum in the definition of $G(k,T+h)$ requires just one comparison since $a(k,T+h) \in \{a(k,T), a(k,T)-1\}$. Given $v(k,T)$ and $a(k,T)$ an optimal trading strategy is straightforwardly implemented using \eqref{eq:cont-dyn}.
 \end{rem}

\section{Extensions}\label{sec:extend}
Using the analysis of Section \ref{sec:toy-model} as a starting point, we now consider several progressively more sophisticated versions of the original model \eqref{eq:v-optim} so as to better express the complexities of real markets.

\subsection{Constrained Trading}\label{sec:constraints}
In this section we consider the modified model whereby $N$ is a compound Poisson process with characteristics $(\lambda,\nu)$ and the agent is constrained to trade only up to the order size $Y_i$. To summarize, we look at the constrained value function
\begin{align*}
\widetilde{v}(k,T) = \inf_{\xi \in \widetilde{\A}_{k}} \E \left[ \sum_{i : \sigma_i \le T} F(\xi_{\sigma_i-}-\xi_{\sigma_i}) +F(\xi_T)\right], \quad k \in \N_+, \,T \in \R_+
\end{align*}
where $\widetilde{\A}_{k}$ is the set of all $\mathbb{F}$ -adapted, integer-valued, positive and decreasing processes whose values change only at the time of jumps of the Poisson process $N$ in such a way that $0 \leq \xi_{\sigma_i-}-\xi_{\sigma_i} \leq Y_i$.
Thus, the model now also includes the distribution $\nu$ of order sizes.
As a first remark, note that we trivially have the bound $\widetilde{v}(k,T) \geq v(k,T)$.

In counterpart to the dynamic programming equation \eqref{eq:dpp}, the constrained value function $\tilde{v}$ is the unique fixed point of the following functional operator $\widetilde{L}$:
\begin{align}
\label{eq:tilde-expectations} \widetilde{L} \widetilde{v} (k,T)&=\E\left[F(k)1_{\{\sigma_1>T\}}+\min_{a \in \{1,2,\ldots,Y_{1} \wedge k\}}\left(F(a)+\widetilde{v}\left(k-a, T-\sigma_1\right)\right)1_{\{\sigma_1 \leq T\}}\right].
\end{align}
The proof of \eqref{eq:tilde-expectations}, as well as of the fact that $\tilde{L}$ has a \emph{unique} fixed point is identical to that of Proposition~\ref{prop:dpp} below and therefore deferred.
Let us now define
\begin{align}\label{eq:wt-a-defn}
\widetilde{a}(k,T) = \argmin_{a \in \{1,\ldots,k\}} \{ \widetilde{v}(k-a,T)+F(a) \}.
\end{align}
The next proposition is analogous to Proposition~\ref{prop:1}. However, it is complicated by the fact the without proving the convexity results regarding $\widetilde{v}(\cdot, T)$ it is not clear
that
\[
\min_{a \in \{1,2,\ldots,Y_{1}\wedge k\}}\left(F(a)+\widetilde{v}\left(k-a, T-\sigma_1\right)\right)=F(\widetilde{a}(k,T) \wedge Y_1)+\widetilde{v}\left(k-(\widetilde{a}(k,T) \wedge Y_1), T-\sigma_1\right).
\]
The above statement implies that an optimal liquidation strategy consists of placing trades of size $\widetilde{a}(k,T)$ and then letting them be filled to the maximum extent by the matching incoming orders.

\begin{prop}\label{prop:constrained}
The following hold:
\begin{enumerate}
\item $k \to \widetilde{v}(k,T)$ is non-decreasing, ``convex", and $\widetilde{v}(k+1,T)-\widetilde{v}(k,T) < F(k+1) -F(k)$ for all $k,T$.

\item $T \to \widetilde{v}(k,T)$ is decreasing and convex.

\item Denote by $\nu[\widetilde{a}(k,T),\infty) = \PP(Y_1 > \widetilde{a}(k,T)).$ Then
\begin{multline}\partial_T \widetilde{v}(k,T) = -\lambda \Big(\widetilde{v}(k,T) - \left[\widetilde{v}(k-\widetilde{a}(k,T),T) + F(\widetilde{a}(k,T)) \right] \nu[\widetilde{a}(k,T),\infty) \\ +\sum_{y=1}^{\widetilde{a}(k,T)}\nu(y)[\widetilde{v}(k-y,T)+F(y)] \Big) <0;
\end{multline} moreover $\partial_T \widetilde{v}(k,T)$ is increasing in $k$.

\item $k \to \widetilde{a}(k, T)$ is non-decreasing and increases by jumps of size 1 only.

\item $T \to \widetilde{a}(k,T)$ is non-increasing and right continuous with $\widetilde{a}(k,0)=\lfloor k/2\rfloor$ and $\lim_{T \to \infty}\widetilde{a}(k,T)=1$. Moreover, its jumps are of size 1. The jumps of $T \to \widetilde{v}(k,T)$ occur at the discontinuity points of $T \to \partial_T^2 \widetilde{v}(k,T)$.

\end{enumerate}
\end{prop}

\begin{proof}
We will first consider an auxiliary control problem in which the agent has to submit her sell orders before seeing the size of the incoming buy orders\footnote{This parallels real markets where once an order is placed, it will be maximally partially filled against any incoming matching order. }. Let us call the corresponding value function by
$\mathcal{V}(k,T)$. Again, a dynamic programming principle implies that this value function is the unique fixed point of an operator $\mathcal{L}$ that is defined by
\[
\mathcal{L} \mathcal{V}(k,T)=\E\left[F(k)1_{\{\sigma_1>T\}}+\left(F(\alpha(k,T) \wedge Y_1)+\mathcal{V}\left(k-(\alpha(k,T) \wedge Y_1), T-\sigma_1\right)\right)1_{\{\sigma_1 \leq T\}}\right],
\]
in which
\[
\alpha(k,T)=\argmin_{a \in \{1,...,k\}}(\mathcal{V}(k-a)+F(a)).
\]
The proofs in Section~\ref{sec:toy-model} now go through to show that the pair $(\mathcal{V}, \alpha)$ satisfies (i)-(v) of Proposition~\ref{prop:constrained}. Now since $\mathcal{V}(\cdot,T)$ is convex, it follows that $\mathcal{V}(k-a,T)+F(a)$ is monotone on the set $\{ a \le \alpha(k,T)\}$ and therefore the action of $\mathcal{L}$ and $\widetilde{L}$ from \eqref{eq:tilde-expectations} against $\mathcal{V}$ is the same. Since $\mathcal{V}$ is a fixed point of $\mathcal{L}$,
\[
\mathcal{V}=\mathcal{L} \mathcal{V}= \widetilde{L} \mathcal{V}.
\]
But $\widetilde{L}$ has a unique fixed point, so that $\tilde{v}(\cdot,\cdot)=\mathcal{V}(\cdot,\cdot)$ and $\alpha(\cdot, \cdot)=\widetilde{a}(\cdot, \cdot)$, and the proof is complete.
\end{proof}



\subsection{Regime Switching Setting}\label{sec:regime-switching}
The model in Section \ref{sec:toy-model} assumed a constant level of trade activity over the full time horizon. However, as practitioners know, real-life order flows experience multiple regime changes. For instance, a common intra-day pattern features high level of activity in the beginning and end of the trading session and a lower trade intensity during midday. Alternatively, markets may experience liquidity crises, whereby order flow abruptly slows down. To capture such stylized features, in this section we assume that $N$ is a regime-switching compound Poisson process, modulated by the market state variable $M$. $M$ represents the market liquidity; namely the
order frequency and order sizes in the order flow book are driven by $M$.

Formally, let $N^{(1)}, \ldots, N^{(m)}$ be $m$ independent compound Poisson processes with
intensities and jump distributions $(\lambda_1, \nu_1), \ldots, (\lambda_m, \nu_m)$. We assume
that $M$ forms an independent finite state Markov chain with state space $E = \{1, 2,
\ldots, m\}$ and infinitesimal generator $Q=(q_{ij})$. Then the observed order flow is given by
\begin{equation}
N_t = \int_0^t \sum_{i \in E} 1_{\{M_s = i\}} dN^{(i)}_s, \quad t \ge 0.
\end{equation}
By construction, the increments of $N$ are independent conditioned on $M$.  Let $v(k,T;i)$ represent the minimal execution costs conditional on $M_0 = i$. Note that the lower and upper bounds derived in Lemmas \ref{lem:lb} and \ref{lem:ub} also bound the value function in the regime switching case. The Hamilton-Jacobi-Bellman equation for the value function is given by the following lemma, also compare with Lemma \ref{lem:v-deriv}.

\begin{lemm}
Let us denote
\[
G(k,T;i):=v(k,T;i)-\min_a [ v(k-a,T;i)+F(a)].
\]
Then derivative of $v$ with respect to its second variable is
\begin{equation}\label{eq:hjb-regime-switching}
\partial_T v(k,T;i) = -\lambda_i G(k,T;i)+\sum_{j \in E\setminus \{i\}} q_{ij}(v(k,T;j)-v(k,T;i)).
\end{equation}
\end{lemm}
\begin{proof}
Denote by $\tau_k$ the $k$-th transition time of $M$.
For $h>0$, let $A=\{\sigma_1>h, \tau_1 > h\}$, $B_N = \{ \sigma_1 < h, \sigma_2>h,\tau_1 >h \}$, $B_{j} = \{\sigma_1 > h, \tau_1 < h, \tau_2 > h, M_{\tau_1}=j\}$, $j \in E \setminus \{i\}$,
 and $C=(A\cup B_N \cup_j B_j)^c$.
By conditional independence of $N$ and $M$ we have that $\P^i(A)=\P(A|M_0=i)=\e^{-(\lambda -q_{ii})h}$, $\P^i(B_N)=\lambda h \e^{-(\lambda-q_{ii}) h}$, $\P^i(B_j) = q_{ij} h \e^{-(\lambda-q_{ii}) h}$  and $\P^i(C)=o(h)$.
Using the dynamic programming principle, we can write
\[
 v(k,T+h; i) = \E^i\left[ v(k,T; i) 1_{A} + (v(k,T;i)-G(k,T; i)) 1_{B_N} + \sum_{j \in E\setminus \{i\}} v(k,T;j)1_{B_j} + X 1_{C}\right],
\]
in which $X$ is a bounded random variable. Taking the limit $h \to 0$ we obtain
\begin{align*}
& \lim_{h\to 0} \frac{ v(k,T+h;i)-v(k,T;i) }{h}  \\ & = \lim_{h\to 0}  \frac{ \E \left[ v(k,T;i)(1_{A \cup B_N \cup_j B_j}) -
G(k,T;i) 1_{B_N} + \sum_{j \in E\setminus \{i\}} (v(k,T;j)-v(k,T;i)1_{B_j}) \right]-v(k,T;i) + o(h)}{h} \\
& = \lim_{h\to 0}  \frac{-\lambda h G(k,T;i) + \sum_{j \in E} q_{ij}h (v(k,T;j)-v(k,T;i))+ o(h)}{h} \\
& = -\lambda G(k,T;i) + \sum_{j \in E\setminus \{i\}} q_{ij}(v(k,T;j)-v(k,T;i)). \qedhere
\end{align*}
\end{proof}









\subsection{Partially Observed Setting}\label{sec:partial-info}
We continue to work with the model of the previous section but now also assume that the market
liquidity variable $M$ is not observed. This is a good proxy for real markets where market
participants do not know the full liquidity state. Instead, agents \emph{infer} current
liquidity based on observed trades. Thus, decreased frequency of trades may point to an
impending liquidity crisis and therefore force agents to place larger trades to avoid being
stuck with an illiquid position.

We shall postulate a Bayesian setting whereby the agent dynamically updates her beliefs about $M$.
Let $D \triangleq \{ \vp \in [0,1]^m \colon \pi_1 + \ldots + \pi_m =1 \}$ be the space of prior
distributions of the Markov process $M$. Let
\begin{equation}
\label{def:P-pi}
\P^{\vp} \{ A \} = \pi_1 \, \mathbb{P} \{A | M_0=1\} + \ldots
+ \pi_m \, \mathbb{P} \{A | M_0=m\}
\end{equation}
for any measurable set $A$. We define the $D$-valued \emph{conditional probability process} $\vP(t) \triangleq \left(
\Pi_1(t), \ldots , \Pi_m(t) \right)$ such that
\begin{align} \label{def:Pi-i}
 \Pi_i(t) = \P^{\vp} \{  M_t =i | \Fc^N_t \}, \quad \text{for $i \in E$, and $t \ge 0$}.
\end{align}
Each component of $\vP$ gives the conditional probability that the current state of $M$ is $\{
i\}$ given the information generated by $N$ until the current time $t$.

The partially-observed execution problem can now be stated as
\begin{equation}\label{eq:v-partial-obs}
v(k,T,\vp) =  \inf_{\xi \in \A^p_k} \E^{\vp} \left[ \sum_{i : \sigma_i \le T} F(\xi(\sigma_i-)-\xi(\sigma_i)) + F(\xi(T)) \right],
\end{equation}
where the minimization is over all $\Fc^N$-adapted admissible controls $\xi$ with $\xi_0=k$. We denote this restricted set of admissible strategies by $\A^p_k$.

With the partially observed setup, the dynamic programming principle for $v(k,T,\vp)$ is no longer trivial. The following proposition establishes such a result using the methods of \cite{BL08,BL08AOR}.

\begin{prop}\label{prop:dpp}
The value function $v$ satisfies the dynamic programming equation $v = Lv$, in which $L$ is the first jump operator given by
\begin{align}
\label{eq:J-expectations} L v (k,T,\vp)&=\E^{\vp}\left[F(k)1_{\{\sigma_1>T\}}+\min_{a \in \{1,2,\ldots,Y_{1}\wedge k\}}\left(F(a)+v\left(k-a, T-\sigma_1, \vP_{\sigma_1}\right)\right)1_{\{\sigma_1 \leq T\}}\right].
\end{align}
In fact, $v$ is the unique fixed point of $L$.
\end{prop}

Before giving the proof of Proposition \ref{prop:dpp}, it is necessary to first understand the behavior of the conditional probability process $\vP$. The sample paths of $\vP$ are obtained as in \cite{BL08}. We briefly summarize the developed theory. First, let
\begin{align}\label{def:I-t}
I(t) \triangleq \int_0^t \sum_{i=1}^m \lambda_i 1_{\{M_s = i \}}\, ds.
\end{align}
By inspection, the expected value of  $\exp(-I(t))$ gives the probability of no events for the next $t$ time units, namely
$\P^{\vp} \{ \sigma_1 > t \} = \E^{\vp}[ \e^{-I(t)}]$. The latter expression is found to be
\citep[Theorem 5.3.2]{Neuts} equal to $\E^{\vp}[ \e^{-I(t)}] = \sum_i m_i(t,\vp)$,
where
\begin{align}\label{def:m}
\vec{m} (t , \vp ) \equiv ( m_1 (t , \vp ), \ldots , m_m (t , \vp ) ) \triangleq \Bigl( \,
\E^{\vp,a} \left[ 1_{\{ M_{t} =1\}} \cdot \e^{ - I(t)}   \right]  , \ldots ,
  \E^{\vp,a} \left[ 1_{\{ M_{t} =m\}} \cdot \e^{ - I(t)}   \right] \, \Bigr)
\end{align}
has the form
\begin{align*}
\vec{m} (t , \vp ) = \vp \cdot \e^{ t( Q - \Lambda) },
\end{align*}
where $\Lambda$ is the $m \times m$ diagonal matrix with $\Lambda_{i,i}  = \lambda_i$. It also
follows that $$\P^{\vp} \left\{ \sigma_1 \in du , M_u =i \right\} = \E^{\vp,a} \left[ \lambda_i
1_{ \{ M_u =i \} } \e^{- I(u)} \right] du =\lambda_i \, m_i(u,\vp) \, du .$$

Consequently, conditional on no arrivals observed on $[t,t+u]$ we obtain using Bayes rule
\begin{align}
\label{eq:semi-group} \Pi_i (t+u) &= \frac{    \P^{\vp}  \{ \sigma_1 > u , M_u =i  \}  }
 {   \P^{\vp}  \{ \sigma_1 > u  \} }\Bigg|_{\vp = \vP(t)} = x_i(u,\vP(t)), \qquad\text{where} \quad\vx(t,\vp) = \frac{m_i(t, \vp)}{\sum_{j \in E} m_j(t, \vp)}.
\end{align}
On the other hand, upon an arrival of order size $Y_\ell$, the conditional probability $\vP$
experiences a jump
\begin{align}\label{eq:jumps-of-vP}
\Pi_i (\sigma_{\ell} ) = \frac{ \lambda_i \nu_i(Y_{\ell}) \Pi_i (\sigma_{\ell} - ) } {
\sum_{j \in E} \lambda_j \nu_j(Y_{\ell})  \Pi_j (\sigma_{\ell} - ) }, \qquad \text{for }\ell
\in \N.
\end{align}

Using the above developments, we are led to define for $i \in E$ the best action operator
\begin{align}
\label{def:S} S_i w(k,T, \vp) \triangleq \sum_{y =1}^\infty \min_{a \le \min(k,y)} \left\{w \left(k-a, T,
\left(\, \frac{ \lambda_1 \nu_1(y) \pi_1 }{ \sum_{j \in E} \lambda_j \nu_j(y) \pi_j }, \ldots,
\frac{ \lambda_m \nu_m(y) \pi_m }{ \sum_{j \in E} \lambda_j \nu_j(y) \pi_j } \right) \right) +
F(a) \right\} \nu_i(y).
\end{align}
Combining \eqref{def:S}-\eqref{eq:semi-group}-\eqref{eq:jumps-of-vP} we see that the action of operator $L$ can be expressed as follows.
\begin{cor}\label{cor:L-expression}
We have
\begin{align}\notag
L v (k,T,\vp)&=\E^{\vp}\left[F(k)1_{\{\sigma_1>T\}}+\min_{a \in \{1,2,\ldots,Y_{1}\wedge k\}}\left(F(a)+v\left(k-a, T-\sigma_1, \vP_{\sigma_1}\right)\right)1_{\{\sigma_1 \leq T\}}\right].
\\ &=  \left(\sum_{i \in E} m_i(T,\vp) \right) \cdot F(k) + \sum_{i
\in E} \int_{0}^{T} m_i(u,\vp) \cdot \lambda_i \cdot S_i v(k, T-u, \vx(u, \vp))
du.
\end{align}
\end{cor}

We now return to the proof of Proposition \ref{prop:dpp}.
\begin{proof}
Let us introduce
\begin{equation}\label{eq:defn-un}
u_0(k,T,\vp)=F(k), \quad u_n (k,T,\vp) \triangleq L u_{n-1}(k,T,\vp), \; n \geq 1.
\end{equation}
Following the logic of the proof of Proposition 3.1 in \cite{BL08}, we can show that
\begin{align}
u_n(k,T,\vp) = v_n(k,T,\vp) \triangleq \inf_{\xi \in \A^p_k} \E^{\vp} \left[ \sum_{i \leq n; \sigma_i \le T} F(\xi_{\sigma_i-}-\xi_{\sigma_i}) +F(\xi_T)\right],
\end{align}
which denotes the value function under the constraint that the agent only trades during the first $n$ orders (and makes zero-trades thereafter until the close $T$).
On the other hand $v_n(k,T,\vp)= v(k,T,\vp)$, for $n \geq k$ since at most $k$ trades are needed to liquidate a position of size $k$. Now, thanks to \eqref{eq:defn-un}
\[
\begin{split}
v(k,T,\vp)&=v_{k+1}(k,T,\vp) = L v_k(k,T,\vp)= Lv(k,T,\vp).
\end{split}
\]

The fact that $v$ is the unique fixed point of $L$, which is an increasing, continuous and concave operator (cf.\ Corollary \ref{cor:L-expression}), follows from standard results in optimal control, see e.g. \cite{zab83} or the proof of Theorem 3.1 in \cite{BL08AOR}.
%
%
%
\end{proof}


In the special case where there are only two liquidity regimes, $E = \{ 1, 2\}$ and identical order size distributions $\nu_1 = \nu_2$ we may obtain an important monotonicity property of the value function.

\begin{lemm} Suppose that $E=\{1,2\}$ and $\lambda_1 > \lambda_2$. Then $\pi \to v(k,T,(\pi,1-\pi))$  is a monotone increasing function.
\end{lemm}
\begin{proof}
With two regimes, we identify the vector $\vp = (\pi, 1-\pi)$ with the scalar $\pi$ and subsequently write $v(k,T,\pi) \equiv v(k,T,(\pi,1-\pi))$, $\PP^{\pi} \equiv \PP^{(\pi,1-\pi)}$, etc.
Observe that the conditional probability of the first arrival time $\E^{\pi}[ \e^{-I(t)}]$ is monotone in $\pi$, and the vector flow $x_1(t,\vp)$ is decreasing in $t$ (as no observed arrivals increase the likelihood of $M$ being in the low-liquidity state 2). Consequently, if $\pi > \pi'$, then $F^{\pi}(t) \ge F^{\pi'}( t)$ for all $t$, where $F^\pi(t) \triangleq \PP^{\pi}(\sigma_1 \le t)$ is the distribution of the first arrival time under the respective measure. Hence, one may construct a probability measure $\tP$ and two random variables $\ts_1 \le \ts_1'$ $\tP$-a.s., such that $\ts_1 \eqd (\sigma_1$ under $\P^{\pi})$ and $\ts_1' \eqd (\sigma_1$ under $\P^{\pi'})$. Moreover, since the jump operator in \eqref{eq:jumps-of-vP} preserves the ordering of $\pi$'s (as does the vector flow $\vx$), it follows that
$$
\Pi^{\pi'}(\ts_1') \le \Pi^{\pi'}(\ts_1) \le \Pi^{\pi}(\ts_1).
$$
Now, conditional on $\ts_1, \ts_1'$, we again have $F^{\Pi^{\pi}(\ts_1)}(t) \ge F^{\Pi^{\pi'}(\ts_1')}(t)$ for all $t$ and therefore we can select interarrival times $\ts_2 \le \ts_2'$ $\tP$-a.s.
with distributions $F_{\ts_2} = F^{\Pi^{\pi}(\ts_1)}$, $F_{\ts_2'} = F^{\Pi^{\pi'}(\ts_1')}$. By the strong Markov property, $\ts_1+\ts_2 \eqd \sigma_2$ (resp.\ $\ts_1'+\ts_2'$) has the same distribution as the second arrival time under $\tP^{\pi}$ (resp.\ $\tP^{\pi'}$). By induction, we construct a measure $\tP$, and  arrivals processes $N^\pi$, $N^{\pi'}$ which satisfy $\tP( N^\pi(t) \ge N^{\pi'}(t) \;\; \forall t) = 1$, while the marginal distributions of $(N^\pi, N^{\pi'})$ are the same as those of $( (N,\PP^{\pi}), (N,\PP^{\pi'}) )$.

We now use the above coupling argument to recursively construct a time-change $\tau(\cdot)$. For $t \le \ts_1'$ define $\tau(t) := (F^{\pi})^{-1}( F^{\pi'}(t) )$. $\tau(\cdot)$ is well-defined since $F^{\pi}(\cdot)$ is strictly increasing for all $\pi$. Moreover, by assumption, $\tau(t) \le t$ for all $t$. Inductively, for $\ts_k' < t \le \ts_{k+1}'$ define
$\tau(t) := (F^{\Pi^{\pi}(\ts_k)})^{-1} ( F^{\Pi^{\pi'}(\ts_k')}(t) )$ (recall that $\ts_k$ and $\ts_k'$ are coupled); then as above we have $\tau(t) \le t$ $\tP$-a.s. and $\tau(\cdot)$ is strictly increasing.
To conclude, observe that the performance of any given control $\xi'$ with respect to $N^{\pi'}$ is the same as the performance of the control $\tilde{\xi}$ with respect to $N^{\pi}$ defined by $\tilde{\xi}(t) := \xi'(\tau^{-1}(t))$, $\tilde{\xi}(t):=\tilde{\xi}(\tau(T))$ for $\tau(T) < t \le T$, since
$\tP( \sigma_k' \le t) = \tP( \sigma_k \le \tau(t))$ for all $k,t$. Thus,
$v_{\xi'}(k,T,\pi') = v_{\tilde{\xi}}(k,T,\pi)$ and since $\xi'$ was arbitrary,
$v(k,T,\pi') \ge v(k,T,\pi)$.

%
%
\end{proof}

\subsection{Continuous Sale Amounts}

A related limiting model is obtained when we allow the sale amounts to be arbitrary real numbers, rather than integers. The corresponding problem becomes
\begin{align}\label{eq:cont-val}
\hat{u}(x,T, \vp) = \inf_{\xi \in {\A^c}_{x}} \E \left[ \sum_{i : \sigma_i \le T} F(\xi_{\sigma_i-}-\xi_{\sigma_i}) +F(\xi_T)\right] \quad x \in \R_+, \,T \in \R_+,
\end{align}
where ${\A}^c_{k} \supseteq \A_{k}$ is now the set of all $\mathbb{F}$ -adapted, non-increasing processes whose values change only at the time of jumps of the Poisson process $N$ with $\xi_0 = x$. The value function when continuous sales are allowed is easier to work with. For example, we can easily derive the following result.
\begin{lemm}
$\hat{u}(x,T,\vp)$ is convex in $x$.
\end{lemm}
\begin{proof}
The proof is immediate once one notes that the set of admissible strategies is convex (which was not true under integer-constraints). Thus, denote by $\xi_1$ (resp. $\xi_2$)
 an $\epsilon$-optimal strategy for $\hat{u}(x_i,t,\vp)$, $i=1,2$. Fix $0 < \lambda < 1$. Then, $\bar{\xi} \triangleq \la \xi_1+(1-\la)\xi_2$ is an admissible strategy for $\hat{u}(\la x_1 + (1-\la)x_2, t,\vp)$ since it will sell $\la x_1$-units using $\xi_1$ and the remaining $(1-\la)x_2$ units using $\xi_2$. Finally,
 \begin{align*}
 \hat{u}(\la x_1 + (1 &-\la)x_2, t,\vp) \le \hat{u}_{\bar{\xi}}(\la x_1 + (1-\la)x_2, t,\vp) \\
 & = \E^{\vp}\left[ \sum_i F( \la (\xi_1(\sigma_i-)-\xi_1(\sigma_i)) +(1-\la)(\xi_2(\sigma_i-)-\xi_2(\sigma_i))) + F( \la \xi_1(T) +(1-\la)\xi_2(T)) \right] \\
 & \le \E^{\vp} \left[ \sum_i \la F( \xi_1(\sigma_i-)-\xi_1(\sigma_i)) +(1-\la)F(\xi_2(\sigma_i-)-\xi_2(\sigma_i)) + \la F(\xi_1(T))+ (1-\la)F(\xi_2(T)) \right] \\
 & = \la \hat{u}(x_1,t,\vp)+\epsilon+(1-\la)\hat{u}(x_2,t,\vp)+\epsilon,
 \end{align*}
 where the penultimate line follows by the convexity of $F(\cdot)$. Since $\epsilon$ was arbitrary the result follows.
\end{proof}

The value function $\hat{u}$ satisfies a scaling property whenever $F$ does. This helps to reduce the dimension of the problem.

\begin{lemm}\label{lem:cont1}
Let us suppose that the depth function $F$ admits the following scaling property, $F(x\beta)/F(\beta) = H(x)$, for some function $H$ and all $\beta > 0$. Then $\hat{u}(x,T,\vp)=H(x) u(T, \vp)$ in which $u$ is the unique solution of
\begin{equation}\label{eq:dpp-ct}
u(T,\vp) = \E^{\vp}\left[H(1)1_{\{\sigma_1>T\}}+\min_{a \in [0,1]} \left(H(a)+H(1-a) \cdot u\left(T-\sigma_1, \vP_{\sigma_1}\right)\right)1_{\{\sigma_1 \leq T\}}\right].
\end{equation}
\end{lemm}

\begin{proof}
Using \eqref{eq:cont-val} and the assumption on $F$ we can see that $\hat{u}(x,T,\vp) = H(x)\hat{u}(1,T,\vp)$ since if $\xi$ is a strategy for $\hat{u}(x,T,\vp)$ then $\xi/x$ is a strategy for $\hat{u}(1,T,\vp)$. With the latter scaling property, the dynamic programming equation \eqref{eq:dpp-ct} is just the counterpart of the original \eqref{eq:dpp}. \end{proof}

Lemma \ref{lem:cont1} leads to the following result which helps us to compute the optimal action directly in the continuous-quantity formulation of the original \eqref{eq:v-optim}.
\begin{cor}\label{cor:cont}
Let us assume that  $F(x) = x^\gamma$ (i.e. $H(x)=x^{\gamma}$), $\gamma>1$. In the framework of the original model \eqref{eq:v-optim}, the function $u$ in Lemma~\ref{lem:cont1} satisfies the following non-linear ordinary differential equation (ODE):
\begin{equation}\label{eq:u-ode}
\partial_T u (T)= \lambda u(T) \left(\frac{1}{[1+u(T)^{1/(\gamma-1)}]^{\gamma-1}}-1\right), \quad u(0)=1.
\end{equation}
Moreover, the optimal action in \eqref{eq:dpp-ct} solves
\begin{equation}\label{eq:a-ode}
\partial_T a (T) = \frac{\lambda}{\gamma-1} a(T) (1-a(T))\left((1-a(T))^{\gamma-1}-1\right)<0, \quad a(0)=1/2,
\end{equation}
and $T \to a(T)$ is convex.
\end{cor}
\begin{proof}
First, the dynamic programming equation \eqref{eq:dpp-ct} leads to the integral equation (note $H(1)=1$)
\[
\begin{split}
u(T) &= e^{-\lambda T} + \int_0^T \min_{a \in [0,1]} \left( a^\gamma +(1-a)^\gamma u(T-s) \right) \lambda e^{-\lambda s} ds
\\&=e^{-\lambda T} \left(1+\int_0^{T} \min_{a \in [0,1]} \left( a^\gamma +(1-a)^\gamma u(s) \right) \lambda e^{\lambda s} ds \right).
\end{split}
\]
The optimal action evidently satisfies
\begin{equation}\label{eq:opt-cont-a-to-ode}
a(T)=\frac{u(T)^{1/(\gamma-1)}}{1+u(T)^{1/(\gamma-1)}}.
\end{equation}
If we let $f(T)=e^{\lambda T} u(T)$, it can be shown that
\[
\partial_T f(T)= \frac{\lambda f(T)}{\left[1+\left(f(T)e^{-\lambda T}\right)^{1/(\gamma-1)}\right]^{\gamma-1}},
\]
from which we can derive the ODE for $u$ in \eqref{eq:u-ode}. Finally, we obtain \eqref{eq:a-ode} for $a$ using \eqref{eq:opt-cont-a-to-ode} and the ODE for $u$.  Since $a(T)\le 1/2$, by inspection the right-hand-side of \eqref{eq:a-ode} is negative and it can also be shown that $\partial^2_T a(T)>0$.
\end{proof}

We find that for a power depth function, Corollary \ref{cor:cont} provides an excellent approximation even for moderate values $k\ge 20$. Thus, when the scaling property of $F$ is satisfied, we obtain a very fast method to compute $v(k,T) \simeq H(k)u(T)$ and $a(k,T)\simeq k \cdot a(T)$ as defined in \eqref{eq:u-ode} and \eqref{eq:a-ode}.

\section{Numerical Illustrations}\label{sec:examples}
In this Section we illustrate the results of our analysis with some computational examples.

\begin{figure}[ht]
\center{\includegraphics[height=3in,width=5in]{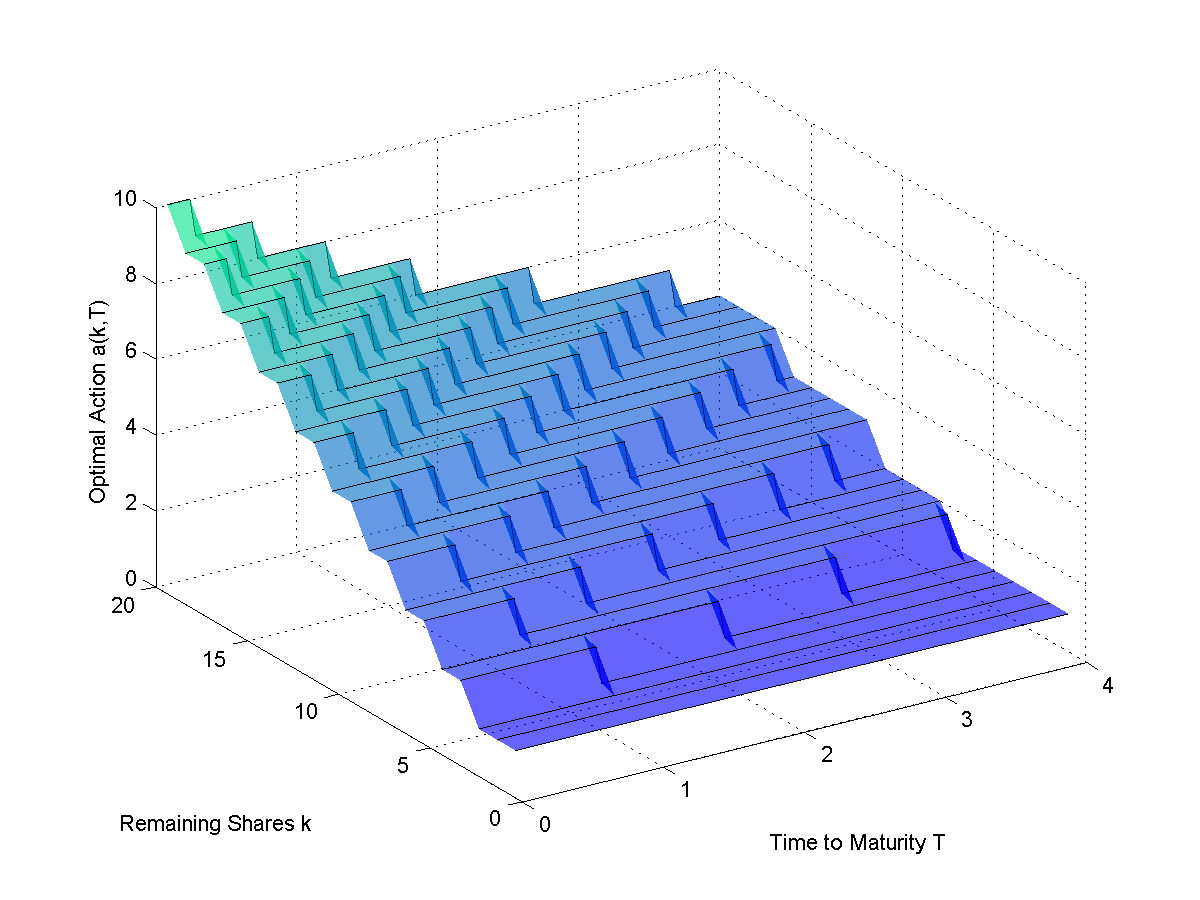}}
\caption{Optimal sale amounts $a(k,T)$ as a function of current holding $k$ and time to maturity $T$. We take $\lambda =1$, $F(a) = a^2/2$ and the model \eqref{eq:v-optim}.
 \label{fig:a-surface}}
\end{figure}

We begin with the base model where we take without loss of generality $\lambda = 1$. We also take a quadratic depth function $F(a) = a^2/2$. Solving for $a(k,T)$ using Remark \ref{rem:computing-v} we obtain Figure \ref{fig:a-surface}. As shown in Lemma \ref{lem:awrtT}, $a(k,\cdot)$ decreases by steps of size 1; at the same time as shown by Corollary \ref{cor:a-diff}, $a(\cdot,T)$ increases by steps of size 1. This surface is used in conjunction with \eqref{eq:cont-dyn} to react to the arrivals of orders in an optimal way.

We then proceed to study the more complex extensions of Section \ref{sec:extend}. Thus, we assume that several liquidity regimes are possible; to be concrete, we fix the liquidity regime-switching model as $M_t \in E = \{ High, Med, Low \} \equiv \{1,2,3\}$ with infinitesimal generator $$ Q =\begin{pmatrix} - 2 & 2 & 0 \\ 1 & -4 & 3 \\ 0 & 2 & -2 \end{pmatrix}.$$
Note that $M$ is recurrent. The intensity of orders is $\lambda(M_t)$ with $\vec{\lambda} = [3, 3, 1]$ and order sizes have
the strictly positive Poisson distributions $\nu_i(y) = \frac{\exp(-\mu_i)}{1-\exp(-\mu i)} \frac{(\mu_i)^y}{y!}$, $y=1,2,\ldots$, with mean sizes $\vec{\mu} = [8, 4, 4]$. The observed order
flow is therefore frequent and of large size in the ``High'' liquidity regime, frequent but small sizes in the ``Med'' regime and infrequent and small order sizes in the
``Low'' regime.

\begin{figure}[ht]
\begin{tabular*}{\textwidth}{lr}
\begin{minipage}{3.2in}
\center{\includegraphics[height=2.4in,width=3.2in]{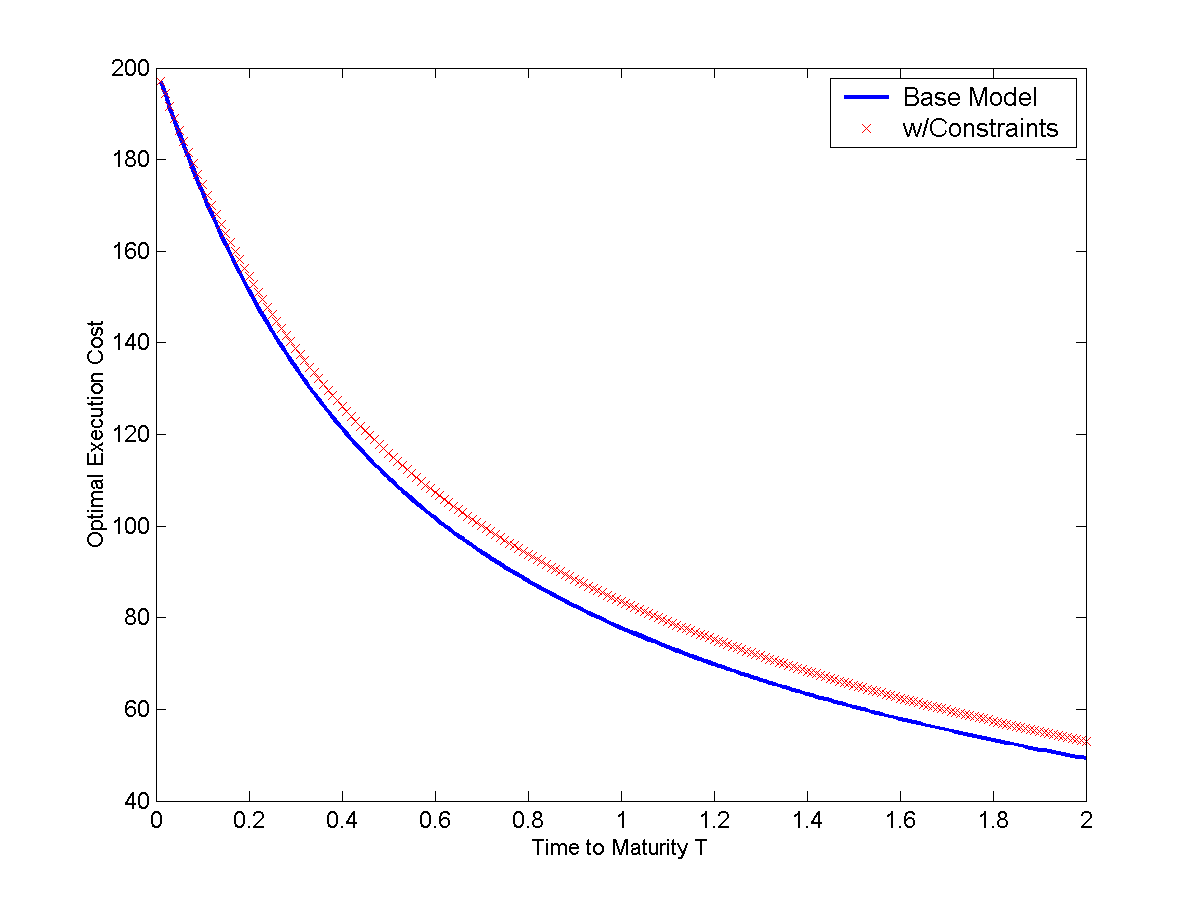}}

\end{minipage}&
\begin{minipage}{3.2in}
\center{\includegraphics[height=2.4in,width=3.2in]{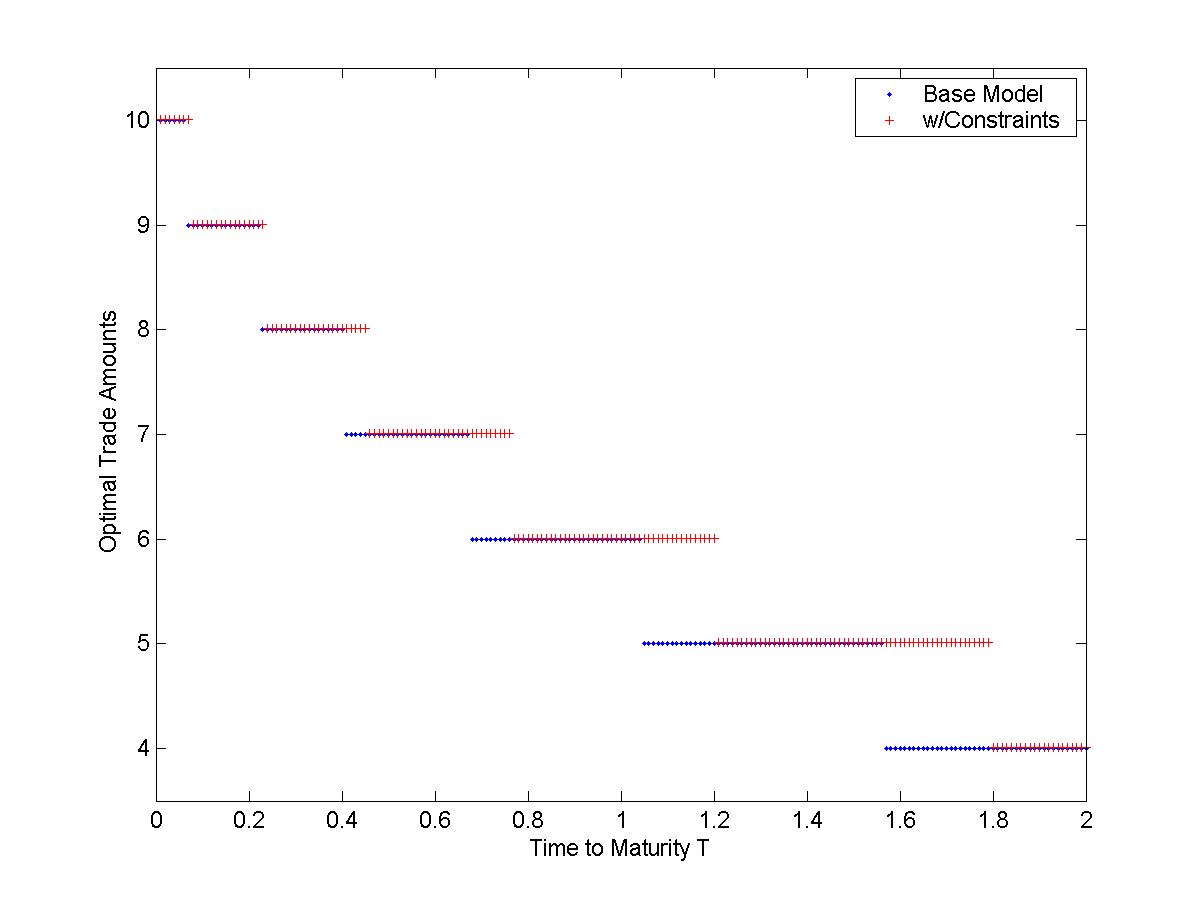}}

\end{minipage}
\end{tabular*}
\caption{The effect of constraints in the regime-switching setting of Section \ref{sec:regime-switching}. Left panel shows the difference between $v(k,T;i)$ and $\tilde{v}(k,T;i)$ for a fixed $i=1$ and $k=20$; right panel plots the difference between $a(k,T; i)$ and $\tilde{a}(k,T;i)$ for same $i=1, k=20$.\label{fig:constraints}}
\end{figure}

In the case of full observations, $v(k,T,i)$ is easily computed by solving the corresponding system of ODE's in \eqref{eq:hjb-regime-switching}. In this context, Figure \ref{fig:constraints} shows the effect of constraints on optimal strategy and optimal execution cost. We observe that constraints play the largest role at medium time horizons, as on long time horizons the agent has plenty of opportunities to trade, while with very short deadlines the convexity of $F$ is the determining factor. Also, as expected the agent responds to constraints by preemptively placing marginally larger orders in the hope they will be filled.

\subsection{Partial Observations}\label{sec:numerical-regimes}
\begin{figure}[h]
\begin{tabular*}{\textwidth}{lcr}
\begin{minipage}{2.2in}
\center{\includegraphics[height=3in,width=2.4in]{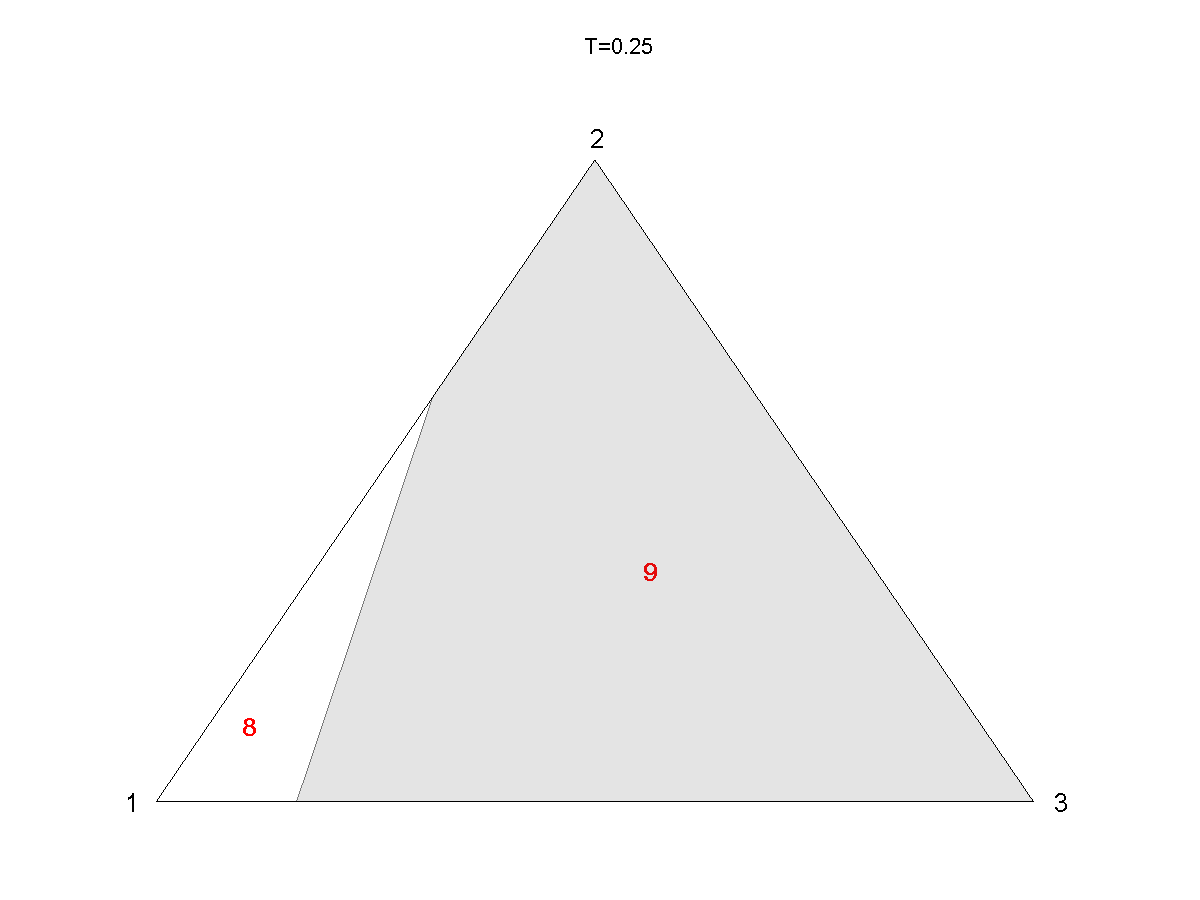}}

\end{minipage} &
\begin{minipage}{2.2in}
\center{\includegraphics[height=3in,width=2.4in]{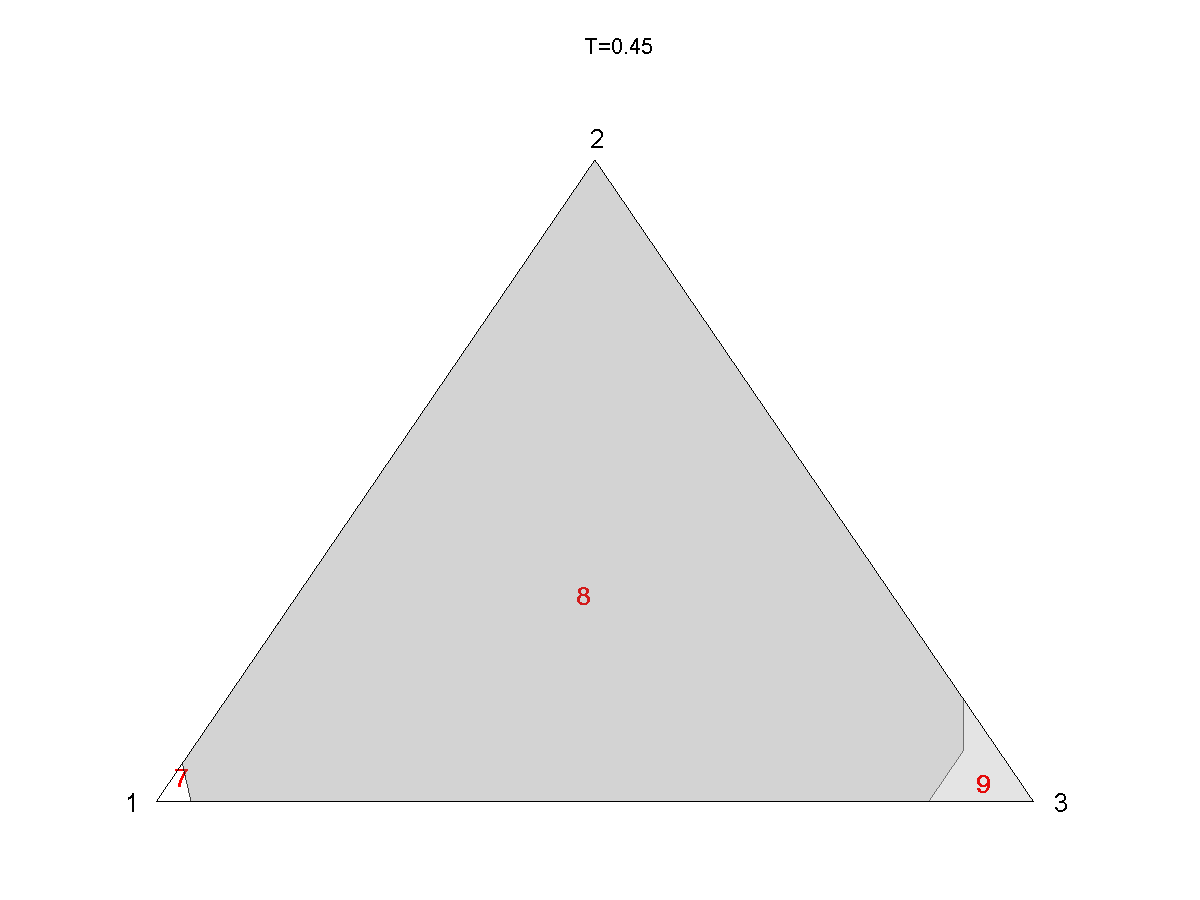}}

\end{minipage} &
\begin{minipage}{2.2in}
\center{\includegraphics[height=3in,width=2.4in]{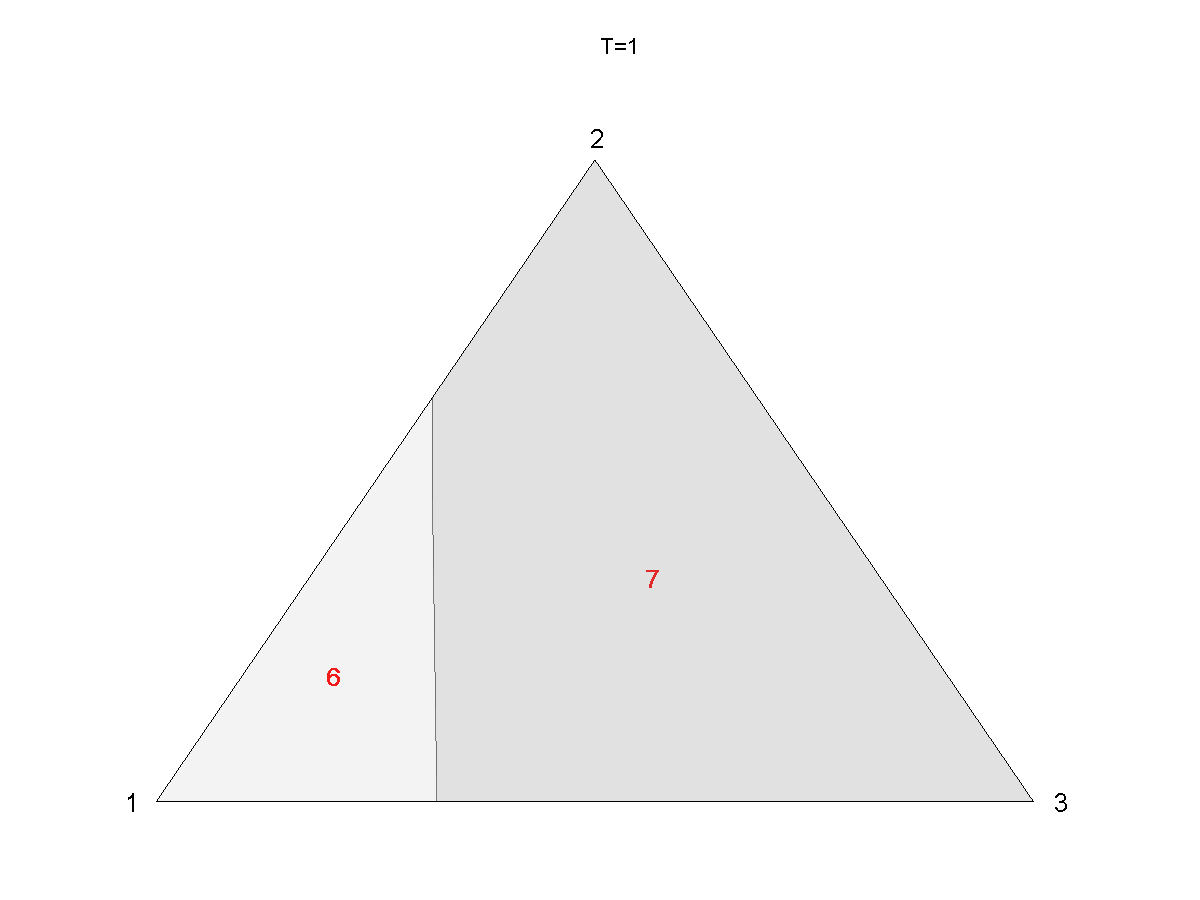}}

\end{minipage}

\end{tabular*}
\caption{Optimal Sale Amounts $a(k,T,\vp)$ for different times to maturity and initial holding of $k=20$ shares for the model of Section \ref{sec:numerical-regimes}. Left panel: $T=0.25$; middle panel: $T=0.45$; right panel: $T=1$. The triangular regions represent the simplex $D=\{ \pi_1+\pi_2 + \pi_3 \le 1, \pi_i \ge 0 \}$ of agent's beliefs. \label{fig:ex0}}
\end{figure}

With the partially-observed setting of Section \ref{sec:partial-info}, the strategies are more complex, as they now depend on the dynamic beliefs $\vP(\cdot)$. Numerically, we compute $v(k,T,\vp)$ and $a(k,T,\vp)$ by solving \eqref{eq:J-expectations} on a discrete mesh approximation of $D=\{ \pi_1+\pi_2 + \pi_3 \le 1, \pi_i \ge 0 \}$ and a discrete time grid with $\Delta t = 0.01$. The action of operator $S_i$ in \eqref{def:S} is obtained by a linear interpolation.
Figure \ref{fig:ex0} shows optimal trading amounts $a(k,T,\cdot)$
for several different horizons $T$ and initial holding of $k=20$ shares. As
expected, as more time is available till the close, optimal order size decreases. We also
see that the beliefs of the trader play an important role; in particular when the likelihood
of being in the ``Low'' frequency regime is large (bottom right corner), the trader will seek to
place larger trades.



To compare the different models of Section \ref{sec:extend}, Table \ref{table:bounds} presents a summary of the various value functions. Namely, we compare the effect of partial observations, and also of constraints. Finally, we also show the accuracy of upper and lower bounds of Lemmas \ref{lem:lb} and \ref{lem:ub} for this case. We see that these bounds are quite tight (relative difference of about 10-15\%) and can be used to give a quick idea about $v$. The bounds are easily computed via a Monte Carlo simulation: one first simulates paths of the continuous-time Markov chain $M$ and then conditional on such a path simulates $N(T)$ using the fact that if $M_s = j$ for $s\in[T_1, T_2]$ then $N(T_2)-N(T_1) \sim Poisson( \la_j (T_2-T_1))$.

The comparison between e.g.\ $v(k,T; 1)$ and $v(k,T, (1,0,0))$ is justified since in both cases the initial system state is the same (namely $M_0 = 1$ $\PP$-a.s.) and therefore the distribution of possible $N$-realizations is identical. This is also the reason why the lower and upper bounds are the same for the fully observed and partially observed models. Thus, $v(k,T, (1,0,0))-v(k,T;1)$ directly measures the effect of partial information on the optimal execution cost.
We find that in the unconstrained case, the effect of partial observations is mild and on the order of 1-2\%. In the given example it is highest in regime 2, which is the ``junction point'' between the favorable ``High'' liquidity regime 1 and the ``Low''-liquidity regime 3. The addition of constraints accentuates the information loss from not observing $M$ since knowledge of $M$ becomes more valuable. Thus, the differences between the partial- and full-observation models are now on the order of 4-5\% in Table \ref{table:bounds}.
Since the formulas in Lemmas \ref{lem:ub} and \ref{lem:lb} are for the base case without constraints, the constrained value functions $\tilde{v}$ are typically larger than the upper bound $\bar{v}$. One could compute an adjusted $\bar{v}$ that takes into account constraints, but no simple formulas like in Lemma \ref{lem:ub} appear to be forthcoming.

\begin{table}
$$ \begin{array}{|c|cccc|}\hline
\multicolumn{5}{|c|}{ \text{Fully Observed Regime Switching}} \\ \hline
  \text{Initial regime } i & \underline{v}(k,T;i)  &  v(k,T;i) & \tilde{v}(k,T;i) & \bar{v}(k,T;i) \\ \hline
1 &  73.16 & 77.80 & 83.54 & 83.31 \\
2 &  84.26 & 88.36 & 98.97 & 93.50 \\
3 &  98.94 & 102.22 & 114.25 & 107.11 \\ \hline \hline
\multicolumn{5}{|c|}{ \text{Partially Observed Regime Switching}} \\ \hline
  \vP_0=\vp & \underline{v}(k,T,\vp)  &  v(k,T,\vp) & \tilde{v}(k,T,\vp) & \bar{v}(k,T,\vp) \\ \hline
(1,0,0) &  73.16 & 78.70 & 86.20 & 83.31 \\
(0,1,0) &  84.26 & 90.49 & 103.14  & 93.50 \\
(0,0,1) &  98.94 & 103.03 & 119.05 & 107.11 \\
(1/3, 1/3, 1/3) & 85.46 & 89.21 & 102.73 & 94.64 \\
\hline
\end{array}$$
\caption{We consider the regime-switching case with $T=1$, $k=20$, $F(a) = a^2/2$. The lower bounds $\underline{v}$ are computed using Lemma \ref{lem:lb} and the upper bound $\bar{v}$ is computed using Lemma \ref{lem:ub}. Note that these bounds are the same for fully-observed and partially-observed settings. We also compare the constrained $\tilde{v}$ to the basic $v$. \label{table:bounds}}
\end{table}


\section{Conclusion}\label{sec:conclusion}
In this paper we have proposed a new model for studying the optimal trade execution problem in financial markets. Our model is directly based on a discrete order flow and therefore is specially suited to capture the features of trading in dark pools where orders are executed only when matched with a crossing counterparty.

To simplify our presentation, our analysis assumed a simple compound Poisson representation of the order flow. However, the obtained dynamic programming equations and most of the stylized properties of the value function and optimal strategy are expected to hold in much more general setups. These could include time-dependent parameters (such as price impact, order intensity and size distribution) or further constraints on optimal strategy.


Realistic dark pool trading involves simultaneous execution on \emph{several} exchanges. In particular, the trader will place trades both in the dark pool and on the regular limit order book in order to optimize the trade-off between liquidity, minimal price impact and information content (dark pool prices are often delayed compared to the limit book). In the case where the order flows of different exchanges are independent, the problem still fits into our framework, since superposition of independent Poisson processes is another Poisson process. The only modification is that orders will now carry the tag of the associated exchange and therefore the depth function $F$ will depend on order type. More complicated multiple-venue problems can be addressed by considering a multi-dimensional version of our model and will be taken up in future work.

\subsection*{Acknowledgment}
This work was initiated at the NSF-CBMS Regional Conference on Convex Duality Method in Mathematical Finance at UC Santa Barbara. We are grateful to the organizers for their hospitality. We also thank Alexander Schied for his stimulating talk that provided the original impetus for our analysis.

\bibliography{scarce-sales}
\bibliographystyle{abbrvnat}

\end{document}